\documentclass[fleqn,usenatbib]{mnras}

\usepackage[T1]{fontenc}
\usepackage{ae,aecompl}

\usepackage{graphicx}	% Including figure files
\usepackage{amsmath}	% Advanced maths commands
\usepackage{amssymb}	% Extra maths symbols
\usepackage{color}

\title[Tides in misaligned systems]{Tidal interactions in spin-orbit misaligned systems}

\author
  [Y. Lin and G. I. Ogilvie]{Yufeng Lin and Gordon I. Ogilvie\\
  Department of Applied Mathematics and Theoretical Physics, University of Cambridge, Centre for Mathematical Sciences,\\ Wilberforce Road, Cambridge CB3 0WA
 }
\date{Accepted XXX. Received YYY; in original form ZZZ}

\pubyear{2016}

\begin{document}
\label{firstpage}
\pagerange{\pageref{firstpage}--\pageref{lastpage}}
\maketitle

\begin{abstract}
Spin-orbit misalignments have been detected in exoplanetary systems and binary star systems. Tidal interactions may have played an important role in the evolution of the spin-orbit angle. In this study, we investigate the tidal interactions in spin-orbit misaligned systems. In particular, we focus on the tidal response of a rotating fluid body to the obliquity tide, which may be important for the evolution of the spin-orbit angle but hardly affects the orbital evolution. The obliquity tide also provides a torque for the mutual precession of the spin and orbital axes around the total angular momentum vector, which has not yet been considered in previous studies on the tidal interactions. In this paper, we first formulate a set of linearized equations describing the  tidal response in spin-orbit misaligned systems, taking into account the precessional motion. Numerical solutions in a homogeneous fluid and in a polytrope of index 1 show that dissipative inertial waves can be excited on top of precession by the obliquity tide in the presence of a rigid core. The tidal quality factor associated with the obliquity tide $Q'_{210}$ can be several orders of magnitude smaller than those associated with other tidal components if their frequencies fall outside the frequency range of inertial waves. Therefore, it is possible that the spin-orbit misalignment undergoes much more rapid decay than the orbital decay in hot Jupiter systems owing to the enhanced dissipation of the obliquity tide.
\end{abstract}


%%%%%%%%%%%%%%%%%%%%%%%%%%%%%%%%%%%%%%%%%%%%%%%%%%

%%%%%%%%%%%%%%%%% BODY OF PAPER %%%%%%%%%%%%%%%%%%

\section{Introduction}
In our solar system, the orbits of all the planets are nearly coplanar with the the Sun's equatorial plane. However, observations of exoplanetary systems have revealed that orbits of exoplanets can be inclined with respect to the equators of their host stars \citep{Winn2015ARAA}, which is known as spin-orbit misalignment. The stellar obliquity, i.e. the angle between the stellar spin axis and the orbital normal, can be determined using the so-called Rossiter-McLaughlin effect for transiting systems \citep{Queloz2000AA}. Among the exoplanets for which the obliquity has been measured, around a third of them show significant spin-orbit misalignments \citep{Barnes2013ApJ}. Spin-orbit misalignments have also been detected in binary star systems \citep{Albrecht2009Nature,Albrecht2014ApJ}. An extreme case of the spin-orbit misalignment in binary star systems is DI Herculis, in which the spin axes of both stars are almost within the orbital plane \citep{Albrecht2009Nature}. A good understanding of the distribution and evolution of the spin-orbit misalignment can provide constraints on stellar and planetary evolution theory in general.

In short-period exoplanetary systems and close binary star systems, tidal interactions may have played an important role in the evolution of spin and orbital configurations \citep{Ogilvie2014}. Indeed, tidal dissipation in the host stars has been invoked to interpret the observed stellar obliquities in hot Jupiter systems \citep{Winn2010ApJ, Albrecht2012ApJ}. They argued that cooler stars ($T_{\mathrm{eff}}\leq6250$K) have low obliquities, whereas the hotter stars have a wide range of obliquities. Therefore, they suggested that the spin-orbit misalignments in the hot Jupiter systems are initially random, but the cooler stars are realigned as a consequence of the effective tidal dissipation in their convective envelopes. However, if the tidal dissipation is efficient in damping the obliquity, the alignment is accompanied by orbital decay and the planet would be destroyed \citep{Barker2009MNRAS,Lai2012MNRAS, Ogilvie2014}. To resolve this conundrum, \citet{Lai2012MNRAS} proposed a modified tidal evolution theory, in which the damping of the misalignment can be much faster than the orbital decay. In a misaligned system, one particular component of the tidal potential (the obliquity tide) has frequency $\hat{\omega}=-\Omega_s$ in the rotating frame ($\Omega_s$ is the stellar spin frequency), which lies within the frequency range of inertial waves.
 On the other hand, the tidal forcing governing the orbital decay usually has frequency $\hat{\omega}\gg2\Omega_s$, which is well beyond the spectrum of inertial waves for typical parameters of hot Jupiter systems. \citet{Lai2012MNRAS} suggested that the obliquity tide can excite inertial waves in the convective envelope, which lead to enhanced dissipation and boost the alignment process, but not the orbital decay. Recent studies have shown that inertial waves can be excited by tides taking into account the Coriolis force, providing a promising channel of tidal dissipation in rotating stars and planets \citep{Ogilvie2004ApJ,Ogilvie2007ApJ,Ogilvie2005JFM,Ogilvie2009MNRAS,Ogilvie2013MNRAS,Wu2005ApJ,Goodman2009ApJ,
 Papaloizou2010MNRAS,Rieutord2010}. 
However, the obliquity tide is in resonance with a trivial inertial wave mode, the so-called spin-over mode, which corresponds to an arbitrary tilt of the spin axis of the fluid body. In fact, the spin-over mode forced by the obliquity tide is just a manifestation of the axial precession in spin-orbit misaligned systems, which does not lead to any dissipation. It might be possible that non-trivial inertial waves can be exited by the obliquity tide on top of precession in the fluid body \citep{Ogilvie2014}, but the underlying mechanism remains to be elucidated. Flows driven by precession have been  studied extensively in the context of the Earth, in which dissipative flows in the liquid core can be driven by precession of the solid mantle through the viscous and topographic couplings at the core-mantle boundary \citep[see review by][ and references therein]{Tilgner2015}. However, the fluid response to precession in stars and gaseous planets has not yet been well studied. \citet{Papaloizou1982MNRAS} studied the precession of gaseous stars using a linear perturbation theory by slightly displacing the rotation axis of the star. By considering the damping of the stellar modes, they also estimated the decay rate of the precessional motion,  which is comparable to the decay rate of the eccentricity in close binary star systems. More recently, \citet{Barker2016MNRAS} investigated turbulent flows driven by axial precession in the context of gaseous giant planets using a local box model.  

The main purpose of our study is to examine whether non-trivial inertial waves can be excited by the obliquity tide and to estimate the associated tidal dissipation, if any. In doing so, we study tidal interactions in spin-orbit misaligned systems by taking into account the mutual precession of the spin and orbital axes around the total angular momentum vector. For simplicity, we consider tidal responses in barotropic fluids, which may contain a rigid core, and assume the orbital companion to be a point mass. In section \ref{sec:equation}, we formulate a set of linearized equations describing the tidal responses in spin-orbit misaligned systems using an asymptotic analysis. In order to seek possible solutions in the form of inertial waves, we decompose the linearized equations into \textit{non-wavelike} and \textit{wavelike} parts following  \citet{Ogilvie2013MNRAS}. In section \ref{sec:Numrics}, we present numerical solutions of the tidal response to the obliquity tide in a homogeneous fluid and in a fluid polytrope of index 1. Our results show that localized inertial waves are excited by the obliquity tide in the presence a rigid core. The dissipation rate resulting from inertial waves depends on the size of the rigid core. In section \ref{sec:Diss}, we estimate the tidal quality factor based upon our numerical calculations and discuss potential implications in hot Jupiter systems and in binary star systems. The paper closes with a conclusion in section \ref{sec:Conc}.
  
\section{Linearized equations} \label{sec:equation}
\begin{figure}
\begin{center}
	\includegraphics[width=0.5\textwidth]{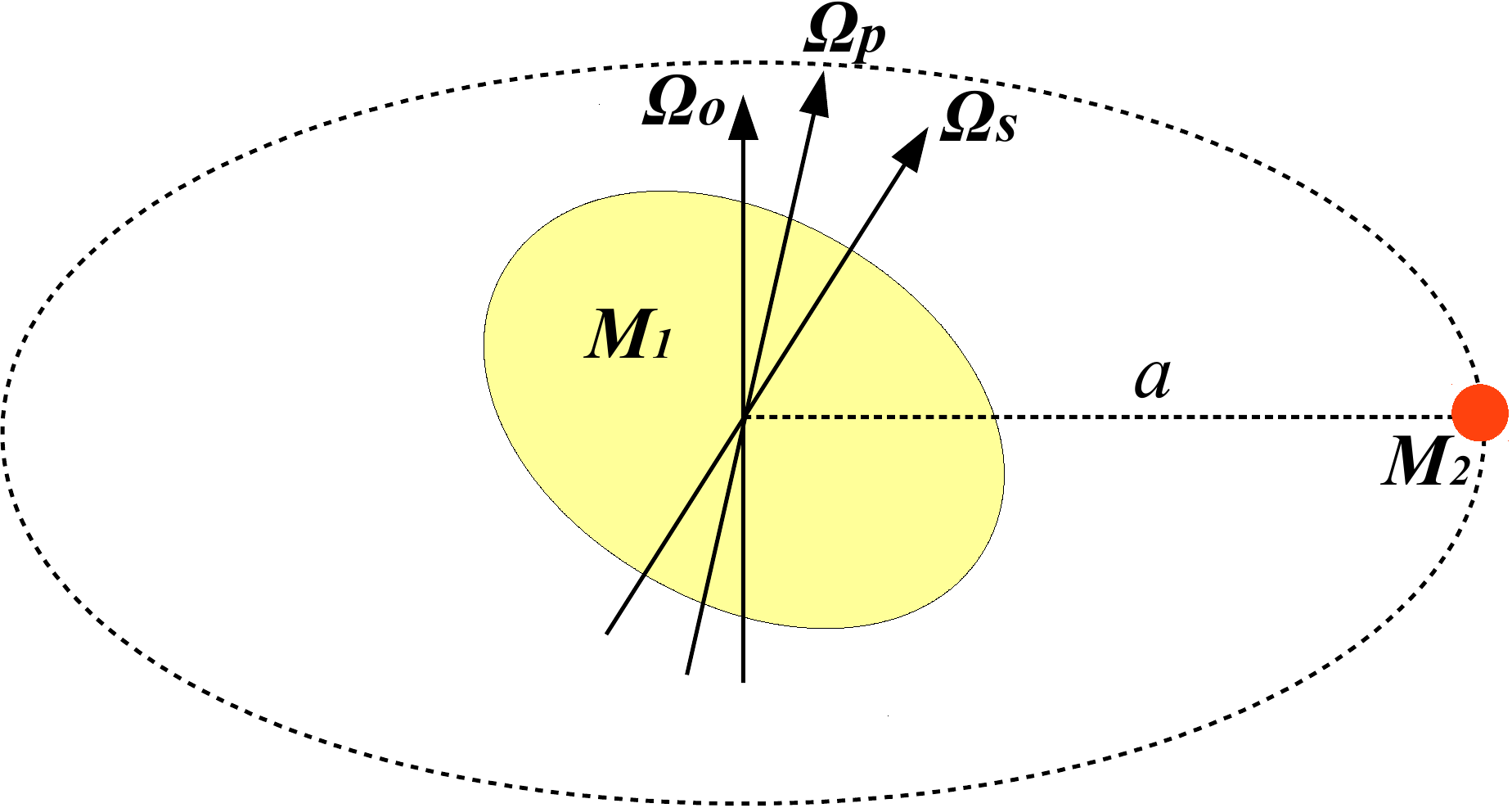}
    \caption{Sketch of the problem.}
    \label{fig:problem}
\end{center}
\end{figure}
We consider two bodies orbiting around the centre of mass of the system, in which the rotation axis  of  body 1 is inclined with respect to the orbital normal, and body 2 is assumed to be a point mass (Fig. \ref{fig:problem}). The inclination angle $i$ is given by
\begin{equation}
\cos i=\bmath{\hat{n}_s} \cdot \bmath{\hat{n}_o},
\end{equation} 
where $\bmath{\hat{n}_s}$ and $\bmath{\hat{n}_o}$ are unit vectors along the spin and orbital angular momentum respectively.  Owing to the tidal torque of body 2 exerted on the equatorial bulge of body 1, the spin axis of body 1 undergoes a forced precession, i.e. a continuous change of the orientation of the spin axis. 
The orbit should also precess because of the conservation of the total angular momentum in the system. In a system in which the spin angular momentum is much smaller than the orbital angular momentum, the spin axis nearly precesses around the orbital normal. In general situations, both the spin vector and the orbital normal vector precess around the total angular momentum vector. In order to take into account the precessional motion, we introduce a frame of reference rotating at $\bmath{\Omega_p}=\Omega_p \bmath{\hat{n}} $ relative to the inertial space, where $\bmath{\hat{n}}$ is the unit vector along the total angular momentum. Although we have introduced such a precessing frame in advance, we do not need to prescribe the precession frequency $\Omega_p$, which will be determined later by the tidal potential. Therefore, precession can still be understood as a part of the tidal response, i.e. the spin-over mode (of finite amplitude) excited by the obliquity tide. In the precessing frame, both $\bmath{\hat{n}_s}$ and $\bmath{\hat{n}_o}$ are stationary (without dissipation). 
The tidal potential $\Psi$ due to body 2 in the precessing frame has the same form as that in the inertial frame without considering precession: 
\begin{equation}
  \Psi= \sum_{l=2}^{\infty} \sum_{m=0}^{l}\sum_{n=-\infty}^{\infty} \frac{GM_2}{a}  \mathcal{A}^m_{l,n}\left(\frac{r}{a}\right)^l Y_l^m(\theta,\phi) \mathrm e^{-\mathrm i n \Omega_o t},
\end{equation}
where $M_2$ is the mass of body 2,  $a$ is the orbital separation, $\Omega_o$ is the mean orbital angular velocity, and $Y_l^m(\theta,\phi)$ is a spherical harmonic in spherical coordinates ($r,\theta,\phi$) centred on body 1 and with $\theta=0$ along the spin axis. The obliquity tide we consider in this study corresponds to the component with $l=2, m=1, n=0$. 

Let body 1 be a self-gravitating barotropic fluid which may contain a rigid core. In the barotropic fluid envelope, the pressure $p$ is a function of the density $\rho$ only.
The momentum equation in the precessing frame can be written as  
\begin{equation}\label{eq:momeq}
 \frac{\partial \bmath{u}}{\partial t}+\bmath u \cdot \nabla \bmath{u}+2\bmath{\Omega_p}\times \bmath{u}=-\nabla h-\nabla \Phi- \nabla \Psi.
\end{equation}
Here $h$ is the specific enthalpy, which is related to the pressure $p$ and the density $\rho$ by \citep{Ogilvie2013MNRAS}
\begin{equation}
\mathrm{d}h=\mathrm{d} p/\rho=V_s^2 \mathrm{d} \rho/\rho, 
\end{equation}
where $V_s=\sqrt{\mathrm{d} p/\mathrm{d}\rho}$ is the sound speed. 

The equation of mass conservation reads
\begin{equation}
 \frac{\partial \rho}{\partial t}+\nabla \cdot (\rho \bmath u)=0.
\end{equation}
The self-gravitational potential $\Phi$ satisfies Poisson's equation,
\begin{equation}
 \nabla^2\Phi=4 \pi G \rho. 
\end{equation}

In order to obtain linearized tidal responses, we introduce two small dimensionless parameters:
\begin{equation}
\epsilon=\frac{\Omega_s}{\omega_d} \quad \mathrm{and} \quad  \beta=\frac{M_2}{M_1}\left(\frac{R}{a}\right)^3,
\end{equation}
where $\omega_d=\sqrt{GM_1/R^3}$ is the dynamical frequency of body 1 with mass $M_1$ and  mean radius $R$. We can see that $\epsilon$ is the ratio of the spin frequency and dynamical frequency of body 1, and $\beta$ is the dimensionless tidal amplitude. We adopt a system of units such that $GM_1=O(1)$ and $R=O(1)$, so the dynamical frequency $\omega_d=O(1)$ and the spin frequency $\Omega_s=O(\epsilon)$. The centrifugal potential is thus of $O(\epsilon^2)$. The tidal potential $\Psi=O(\beta)$ for the $l=2$ components, which implies that the leading-order tidal deformation is of $O(\beta)$. In a low-frequency limit, i.e. when the tidal frequency $\omega=O(\epsilon)$, the velocity perturbation is then of $O(\epsilon \beta)$. The scaling for the precession frequency can be estimated based on the conventional formula derived by calculating the torque applied on a deformable spheroid \citep{Kopal1969ApSS}  
\begin{equation}\label{eq:omgp}
\Omega_p=-\frac{3 GM_2}{2 a^3 \Omega_s} \frac{I_3-I_1}{I_3}\cos i,
\end{equation}
where $I_3$ and $I_1$ are moments of inertia around the spin axis and around an axis in the equatorial plane respectively. The quadrupole moment $I_3-I_1=\frac{k_2}{3} M_1R^2 \Omega_s^2/\omega_d^2$, and $I_3=k_* M_1R^2$, where $k_2$ is the Love number and $k_*$ is the moment of inertia constant. We end up with the scaling of $\Omega_p=O(\epsilon \beta)$, provided that $k_2/k_*=O(1)$.   

 Based upon the above scalings, we may cast the asymptotic expansions in small parameters $\epsilon$ and $\beta$:
  \begin{equation}\label{eq:u_exp}
\bmath u = \epsilon \bmath{u}_1(\bmath r) +\beta \epsilon \bmath u_1'(\bmath r, t)+O(\epsilon^5,\beta \epsilon^3, \beta^2 \epsilon), 
 \end{equation}
 \begin{equation}
\rho  = \rho_0(r)+\epsilon^2 \rho_2 (\bmath r)+\beta \rho_1'(\bmath r,t)+O(\epsilon^4,\beta \epsilon^2,\beta^2), 
 \end{equation}
 \begin{equation}
h   =  h_0(r)+\epsilon^2 h_2 (\bmath r)+\beta h_1'(\bmath r,t)+O(\epsilon^4,\beta \epsilon^2,\beta^2),
\end{equation}
\begin{equation}
\Phi= \Phi_0(r)+\epsilon^2 \Phi_2 (\bmath r)+\beta \Phi_1'(\bmath r,t)+O(\epsilon^4,\beta \epsilon^2,\beta^2),
\end{equation}
where all terms with primes represent the Eulerian perturbations due to the tidal potential.   
For the sake of convenience, we also use the following scalings:
 \begin{equation} \label{eq:omgs1}
 \Omega_s=\epsilon \Omega_{s1}, 
  \end{equation}
  \begin{equation} \label{eq:omgp1}
  \Omega_p = \epsilon \beta \Omega_{p1},
  \end{equation}
\begin{equation} \label{eq:psi1}
 \Psi  =  \beta \Psi_1,
 \end{equation}
such that all quantities with numerical subscripts are of the order of unity.  In equation (\ref{eq:u_exp}),  $\bmath{u}_1=\bmath{\Omega_{s1}}\times \bmath r$ corresponds to a rigid body rotation. 

At leading order,  we obtain the spherical hydrostatic equilibrium:
\begin{equation}
  g=\frac{d \Phi_0}{d r}=-\frac{V_s^2}{\rho_0} \frac{d \rho_0}{d r}.
\end{equation}

At order $\epsilon^2$, we obtain the rotationally deformed hydrostatic equilibrium:
\begin{equation}\label{eq:hydro}
 0=-\nabla (h_2+\Phi_2+\Phi_c),
\end{equation}
where $\Phi_c=\frac{1}{2}|\bmath{\Omega_{s1}}\times \bmath r|^2$ is the centrifugal potential.

At order $\beta$, we have equilibrium tidal perturbations:
\begin{equation}
0=\nabla(h_1'+\Phi_1'+\Psi_1),
\end{equation}
which implies that $h_1'+\Phi_1'+\Psi_1=\mathrm{constant}$. 
Since the tidal potential $\Psi_1$ is a sum of harmonic components, each with zero mean, and the tidal perturbation $(h_1',\Phi_1')$ has a similar property, the constant vanishes. We may write the above equation as
\begin{equation}\label{eq:h2}
h_1'+\Phi_1'+\Psi_1=0.
\end{equation}

From the momentum equation at order $\beta \epsilon^2$, we obtain the linearized velocity perturbations
\begin{equation}\label{eq:u4}
  \frac{\partial \bmath u_1'}{\partial t}+\Omega_{s1}\frac{\partial \bmath u_1'}{\partial \phi}+\bmath{\Omega_{s1}}\times \bmath u_1'+2\bmath{\Omega_{p1}}\times(\bmath{\Omega_{s1}}\times \bmath r)= -\nabla W',
\end{equation}
where $W'$ is associated with perturbations of $O(\beta \epsilon^2)$ in $h$ and $\Phi$. 
To obtain equation (\ref{eq:u4}), we have made use of the following identity:
\begin{equation}
\bmath u_1 \cdot \nabla \bmath u_1'+\bmath u_1' \cdot \nabla \bmath u_1=\Omega_{s1}\frac{\partial \bmath u_1'}{\partial \phi}+\bmath{\Omega_{s1}}\times \bmath u_1',
\end{equation}
where $\bmath{u}_1=\bmath{\Omega_{s1}}\times \bmath r$. The last term on the left-hand side of equation (\ref{eq:u4}) corresponds to $2 \bmath{\Omega_p}\times \bmath{u}_1$  in the momentum equation (\ref{eq:momeq}).

The mass conservation equation at order $\beta \epsilon$ can be written as 
\begin{equation}\label{eq:rho2}
 \frac{\partial \rho_1'}{\partial t}+\Omega_{s1}\frac{\partial \rho_1'}{\partial \phi}+\nabla \cdot(\rho_0 \bmath u_1')=0,
\end{equation}
and the self-gravitational potential perturbation follows from
\begin{equation} \label{eq:phi2}
 \nabla^2 \Phi_1'=4 \pi G \rho_1'.
\end{equation}

Equations (\ref{eq:h2}-\ref{eq:phi2}) govern linear tidal perturbations at leading order.
We can omit the subscripts of 1 in equations (\ref{eq:h2}-\ref{eq:phi2})  by  using
\begin{equation} \label{eq:pert}
\bmath{u'}=\beta \epsilon \bmath u_1',\quad \rho'=\beta \rho_1',\quad h'=\beta h_1', \quad \Phi'=\beta \Phi_1',
\end{equation}
  and equations (\ref{eq:omgs1}-\ref{eq:psi1}). We do not need solutions for equation (\ref{eq:hydro}) at $O(\epsilon^2)$, as far as the tidal perturbations in equation (\ref{eq:pert}) are concerned. Therefore, we can assume that the basic state is spherically symmetric and may contain a spherical rigid core of radius $\eta R$.

We look for solutions of perturbations proportional to $\mathrm{exp}(\mathrm{i} m \phi-\mathrm{i}\omega t)$, which has the same azimuthal wavenumber $m$ and frequency $\omega$ as the tidal forcing.   
Equations (\ref{eq:h2}-\ref{eq:phi2}) can be written in the fluid frame, which is rotating with the angular velocity $\bmath{\Omega_s}$ when viewed from the precessing frame,
\begin{equation}\label{eq:uf}
  \frac{\partial \bmath u'}{\partial t}+2\bmath{\Omega_s}\times \bmath u'= -\nabla W'-(\bmath{\Omega_p}\times \bmath{\Omega_s})\times \bmath r,
\end{equation}
\begin{equation}\label{eq:hf}
h'+\Phi'+\Psi=0.
\end{equation}
\begin{equation}\label{eq:rhof}
 \frac{\partial \rho'}{\partial t}+\nabla \cdot(\rho_0 \bmath u')=0,
\end{equation}
\begin{equation} \label{eq:phif}
 \nabla^2 \Phi'=4 \pi G \rho'.
\end{equation}
In equation (\ref{eq:uf}), we have made use of the following identity:
\begin{equation}
2\bmath{\Omega_p}\times (\bmath{\Omega_s}\times \bmath r)+\nabla[(\bmath{\Omega_p}\times \bmath r)\cdot(\bmath{\Omega_s}\times \bmath r)]=(\bmath{\Omega_p}\times \bmath{\Omega_s})\times \bmath r
\end{equation}
and combined the potential term into $\nabla W'$. 

The boundary conditions are then given as \citep{Ogilvie2013MNRAS}
  \begin{equation}
\bmath{\hat{r}}\cdot \bmath{u'}=0 \quad \mathrm{at} \quad r=\eta R,  
\end{equation}    
   \begin{equation}
\bmath{\hat{r}} \cdot \bmath{u'}=-\frac{\dot{\Phi}'+\dot{\Psi}}{g} \quad \mathrm{at} \quad r= R.  
\end{equation}  
In the fluid frame, the spin angular velocity $\bmath{\Omega_s}=\Omega \bmath{\hat{n_s}}$ is steady and the precessional angular velocity $\bmath{\Omega_p}=\Omega_p \bmath{\hat{n}}$ is time dependent \citep{Lin2015}:
\begin{equation}
\bmath{\hat{n_s}}=\bmath{\hat{z}},
\end{equation}
\begin{equation}
\bmath{\hat{n}}=-\sin \alpha_s [\bmath{\hat{x}} \cos(\Omega_s t)-  \bmath{\hat{y}}\sin(\Omega_s t)]+\cos \alpha_s \bmath{\hat{z}}, 
\end{equation}
where $\alpha_s$ is the angle between the spin angular momentum and total angular momentum,
\begin{equation}
\cos \alpha_s=\bmath{\hat{n_s}}\cdot \bmath{\hat{n}}.
\end{equation}

Equations (\ref{eq:uf}-\ref{eq:phif}) are the same as those in \citet{Ogilvie2013MNRAS} except for an additional precessional forcing $-(\bmath{\Omega_p}\times \bmath{\Omega_s})\times \bmath r$ in the momentum equation (\ref{eq:uf}).   
These equations can be solved using the decomposition of \textit{non-wavelike} and \textit{wavelike} introduced by \citet{Ogilvie2013MNRAS}. 
The non-wavelike part can be determined by
\begin{equation} \label{eq:nw_u}
  \bmath u_{nw}=-\nabla \dot{X},
\end{equation}
\begin{equation} \label{eq:nw_X}
 \nabla \cdot (\rho_0 \nabla X)=-\frac{\rho_0}{V_s^2}(\Phi'+\Psi),
\end{equation}
\begin{equation} \label{eq:nw_Phi}
 \nabla^2\Phi'+\frac{4 \pi G \rho_0}{V_s^2}(\Phi'+\Psi)=0.
\end{equation}

The wavelike part satisfies the following equations:
\begin{equation} \label{eq:wave_NS}
  \frac{\partial \bmath u_w}{\partial t}+2\bmath{\Omega_s}\times \bmath u_w=\nabla W_w +\bmath{f_p}+\bmath{f_{nw}},
\end{equation}
\begin{equation} \label{eq:wave_mass}
 \nabla\cdot(\rho_0 \bmath u_w)=0,
\end{equation}
where
\begin{equation}
\bmath{f_p}=-(\bmath{\Omega_p}\times \bmath{\Omega_s})\times \bmath r,
\end{equation}
\begin{equation}
\bmath{f_{nw}}=-2\bmath{\Omega_s}\times \bmath{u_{nw}}.
\end{equation}
The boundary conditions are also decomposed as
\begin{equation}
\bmath{\hat{r}}\cdot\bmath u_{nw}=0 \quad \mathrm{at} \quad r=\eta R,
\end{equation}
\begin{equation}
\bmath{\hat{r}}\cdot\bmath u_{nw}=-\frac{\dot{\Phi'}+\dot{\Psi}}{g} \quad \mathrm{at} \quad r=R,
\end{equation}
\begin{equation}
\bmath{\hat{r}}\cdot\bmath u_{w}=0 \quad \mathrm{at} \quad r=\eta R \quad \mathrm{and} \quad r=R.
\end{equation}

The non-wavelike part is an instantaneous response to the tidal potential. For a given tidal component
\begin{equation}
\Psi_{lmn}=A_{lmn}r^l Y_l^m(\theta,\phi) \mathrm{e}^{-\mathrm{i} \hat{\omega} t},
\end{equation}
the corresponding $\Phi'$ and $X$ admit solutions of the form 
\begin{equation}
\Phi'=\Phi'_{lmn}(r)Y_l^m(\theta,\phi) \mathrm{e}^{-\mathrm{i}\hat{\omega} t},
\end{equation}
\begin{equation}
X=X_{lmn}(r)Y_l^m(\theta,\phi) \mathrm{e}^{-\mathrm{i}\hat{\omega} t},
\end{equation}
where $\hat{\omega}=n\Omega_o-m\Omega_s$ is the tidal frequency in the fluid frame. Equations~ (\ref{eq:nw_X}-\ref{eq:nw_Phi}) then reduce to ODEs \citep{Ogilvie2013MNRAS}:
\begin{multline}\label{eq:nw_Phi_ODE}
\frac{1}{r^2}\frac{\mathrm{d}}{\mathrm{d}r}\left(r^2\frac{\mathrm{d} \Phi'_{lmn}(r)}{\mathrm{d}r}\right)-\frac{l(l+1)}{r^2}\Phi'_{lmn}(r)\\
+\frac{4\pi G \rho_0}{V_s^2}\left(\Phi'_{lmn}(r)+A_{lmn}r^l\right)=0,
\end{multline}
\begin{multline}\label{eq:nw_X_ODE}
\frac{1}{r^2}\frac{\mathrm{d}}{\mathrm{d}r}\left(r^2\rho_0\frac{\mathrm{d} X_{lmn}(r)}{\mathrm{d}r}\right)-\frac{l(l+1)}{r^2}X_{lmn}(r) \\
+\frac{\rho_0}{V_s^2}\left(\Phi'_{lmn}(r)+A_{lmn}r^l\right)=0.
\end{multline}
By solving these ODEs, we can obtain the solution for the non-wavelike part and the Coriolis force due to the non-wavelike  velocity $\bmath{u_{nw}}$.

The wavelike solutions are inertial waves forced by the precessional forcing $\bmath{f_p}$ and the Coriolis force $\bmath{f_{nw}}$ due to the non-wavelike  velocity. As we have mentioned before, there is a formal resonance between the tidal component $\Psi_{210}$ and the spin-over mode, which effectively leads to precession of the spin axis. This formal resonance can be removed by using the solvability condition
\begin{equation}
  \int_V \rho_0 \left( \bmath f_{p}+\bmath f_{nw} \right) \cdot  \bmath u_{so}^* dV=0,
\end{equation}
where $\bmath{u_{so}}$ is the spin-over mode (of arbitrary amplitude) in the fluid frame 
\begin{equation}
  \bmath u_{so}= r \mathrm{e}^{\mathrm{i}( \Omega_s t+\phi)}\bmath{\hat{\theta}} +\mathrm{i}r m\cos \theta \,\mathrm{e}^{\mathrm{i}( \Omega_s t+\phi)}\bmath{\hat{\phi}}. 
\end{equation}
The integral is carried out over the fluid volume. The solvability condition applied to the obliquity tide $\Psi_{210}$  determines the precession frequency $\Omega_p$, at which the tidal torque due to the obliquity tide is exactly balanced by the precessional motion. In other words, we have chosen a frame of reference in which the spin-over mode is not allowed to be excited any more. This also requires the wavelike solution satisfy to 
\begin{equation}\label{eq:constraint}
\int_V \rho_0 \bmath u_{w} \cdot  \bmath u_{so}^* dV=0.
\end{equation}

Note that the solvability condition applied to other tidal components gives $\Omega_p=0$, which implies that other tidal components do not affect the precession frequency.

\section{Numerical solutions} \label{sec:Numrics}
In this section, we seek numerical solutions of the linearized equations obtained in section \ref{sec:equation} and consider only the obliquity tide
\begin{equation}
\Psi_{210}=A_{210}r^2Y_2^1(\theta,\phi) \mathrm{e}^{\mathrm{i}\Omega_s t},
\end{equation}
 where  $A_{210}=\frac{GM_2}{a^3}\sqrt{\frac{6 \pi}{5}} \sin i \cos i$ in a misaligned system   \citep{Barker2009MNRAS,Lai2012MNRAS}.
   This tidal component always has frequency $\hat{\omega}=-\Omega_s$ in the fluid frame regardless of the orbital frequency. The obliquity tide provides a torque for the precessional motion. If dissipative inertial waves can be excited on top of precession by the obliquity tide, the resulting dissipation would play an important role in the evolution of the spin-orbit misalignment but hardly affect the orbital evolution. Bearing in mind the main motivation to see whether non-trivial inertial waves can be excited by the obliquity tide, we will focus only on the obliquity tide in this section. 
\subsection{Numerical method}
 The non-wavelike solution can be obtained by solving ODEs (\ref{eq:nw_Phi_ODE}-\ref{eq:nw_X_ODE}) using a Chebyshev collocation method. For the wavelike equations, we need to introduce a viscous force for the dissipation in equation (\ref{eq:wave_NS}). In this study, we consider the dissipative force in the form of $\nu \nabla^2 \bmath {u_w}$, where  $\nu=\mu/\rho_0$ is the kinematic viscosity. For numerical convenience, we assume the kinematic viscosity $\nu$ is uniform. The viscous effect is measured by a dimensionless parameter, the Ekman number
\begin{equation}
E=\frac{\nu}{\Omega_sR^2},
\end{equation} 
which is the typical ratio between the viscous force and the Coriolis force. The time-averaged dissipation rate resulting from the wavelike velocity is 
\begin{equation}\label{eq:dissrate}
D=-\frac{1}{2}\mathrm{Re}\int_V \rho_0 \nu  \bmath {u_w}^*\cdot \nabla^2 \bmath {u_w} \mathrm{d}V,
\end{equation}
where the star symbol denotes the complex conjugate. 
 
 The equations governing the wavelike part are numerically solved using a pseudo-spectral method as described in \citet{Ogilvie2004ApJ} and \citet{Ogilvie2009MNRAS}.  We use a spheroid-toroidal decomposition 
\begin{equation} \label{eq:SphTor_ur}
 u_r=\sum_n a_n(r)Y_n^m(\theta,\phi),
\end{equation}
\begin{equation}\label{eq:SphTor_ut}
 u_\theta=r\sum_n \left[b_n(r)\frac{\partial}{\partial \theta}+c_n(r)\frac{\mathrm{i}m}{\sin \theta}\right]Y_n^m(\theta,\phi),
\end{equation}
\begin{equation}\label{eq:SphTor_up}
 u_\phi=r\sum_n \left[b_n(r)\frac{\mathrm{i}m}{\sin \theta}-c_n(r)\frac{\partial}{\partial \theta}\right]Y_n^m(\theta,\phi),
\end{equation}
\begin{equation}\label{eq:SphTor_W}
 W=\sum_n W_n(r)Y_n^m(\theta,\phi).
\end{equation}   
The wavelike equations (\ref{eq:wave_NS}-\ref{eq:wave_mass}) can be projected onto the spherical harmonics. The projected equations are given in Appendix \ref{app:projection}. These equations are truncated at spherical harmonic degree $L$ numerically.  

For the radial dependence, we use Chebyshev collocations on $N+1$ Gauss-Lobatto nodes. We use the stress-free boundary condition on the free surface
\begin{equation}
  a_n=\frac{d b_n}{d r}=\frac{d c_n}{d r}=0.
\end{equation}
On the inner core boundary, we use either the stress-free boundary condition or the no-slip boundary condition
\begin{equation}
 a_n=b_n=c_n=0. 
\end{equation}
The additional constraint in equation (\ref{eq:constraint}) only involves $c_1(r)$ because of the orthogonality of the spherical harmonics and can be written as
\begin{equation}
\int_{\eta R}^R \rho_0(r) c_1(r)r^4 \mathrm{d}r=0.
\end{equation}
The integral can be evaluated using a Chebyshev quadrature formula, which involves the function values at the Gauss-Lobatto nodes. 
\subsection{In a homogeneous fluid}
\begin{figure*}
\includegraphics[width=0.33 \textwidth,clip,trim=2cm 1.5cm 1cm 0cm]{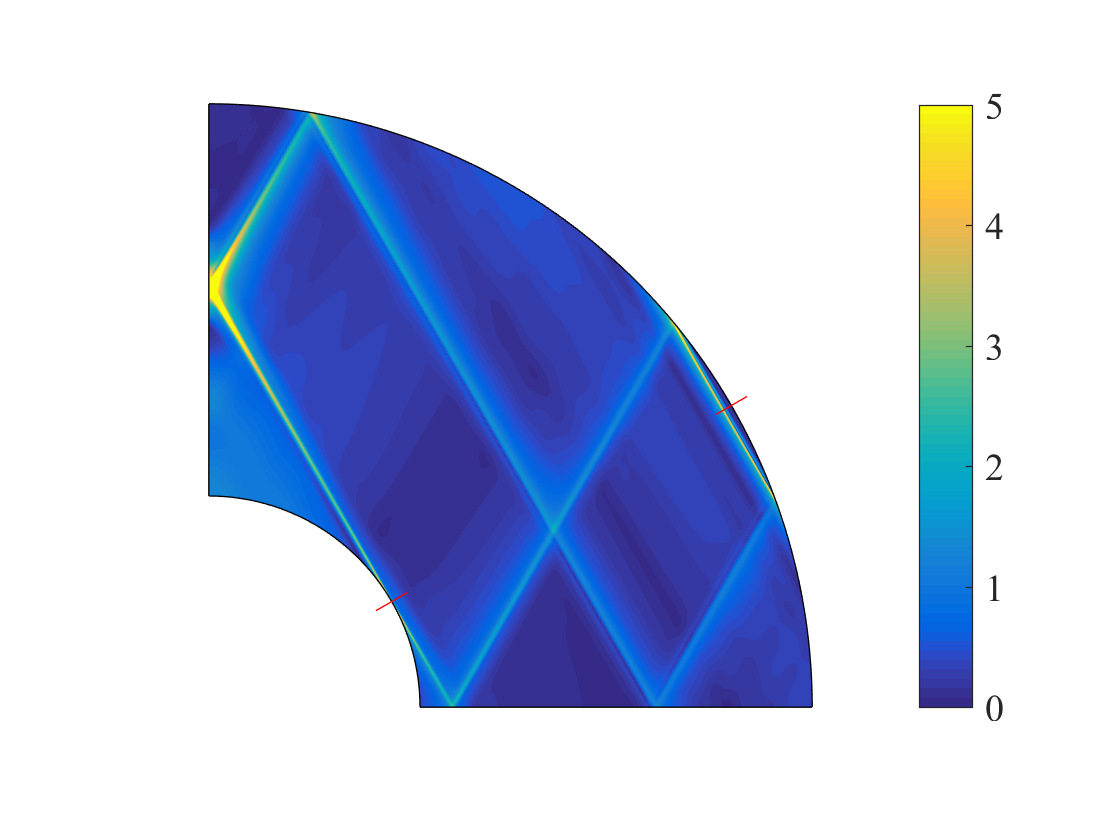}
\includegraphics[width=0.33 \textwidth,clip,trim=2cm 1.5cm 1cmcm 0cm]{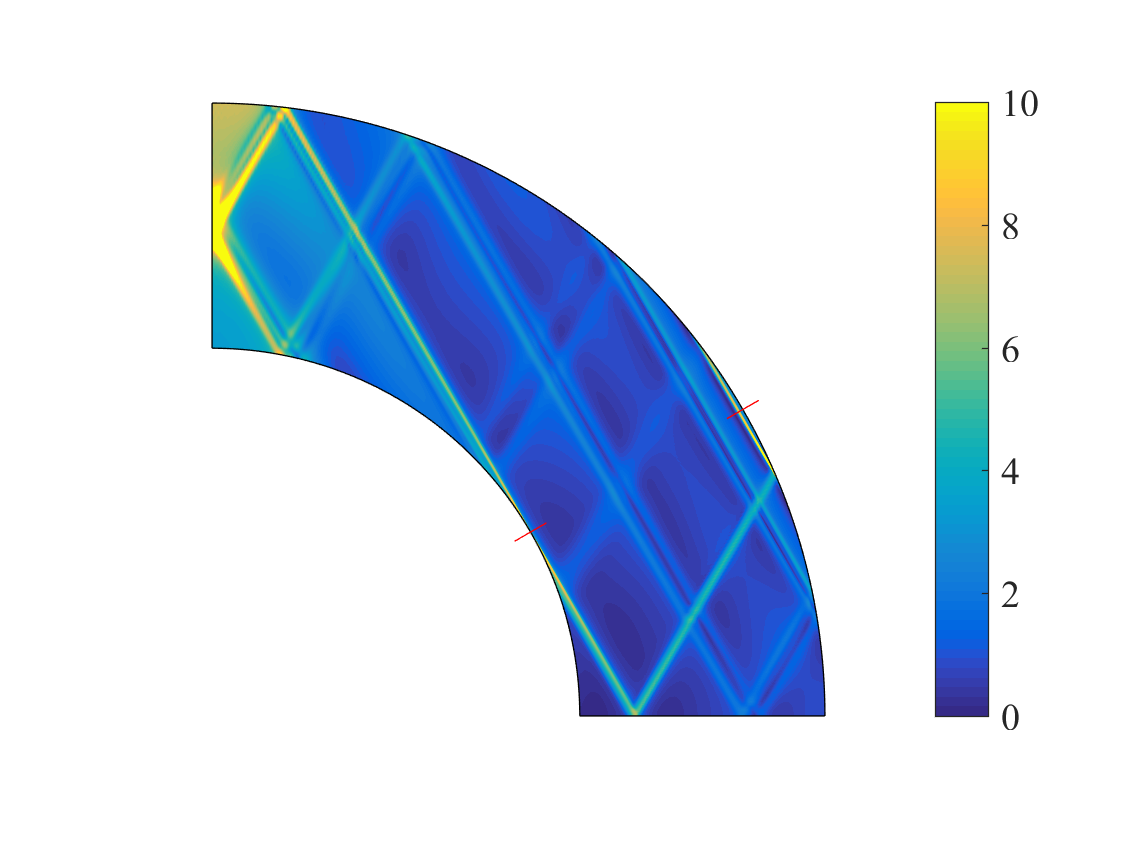}
\includegraphics[width=0.33 \textwidth,clip,trim=2cm 1.5cm 1cm 0cm]{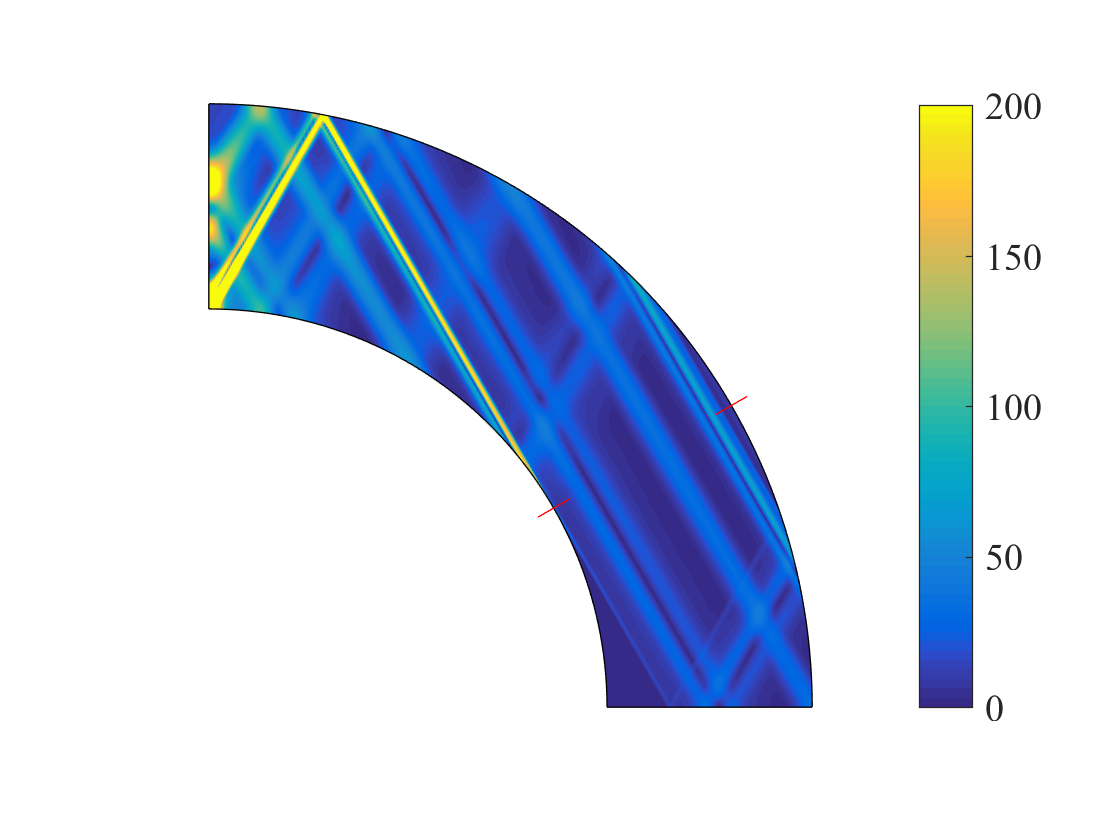} \\
(a) \hfill (b) \hfill(c) \\
\includegraphics[width=0.33 \textwidth,clip,trim=2cm 1.5cm 0.8cm 0cm]{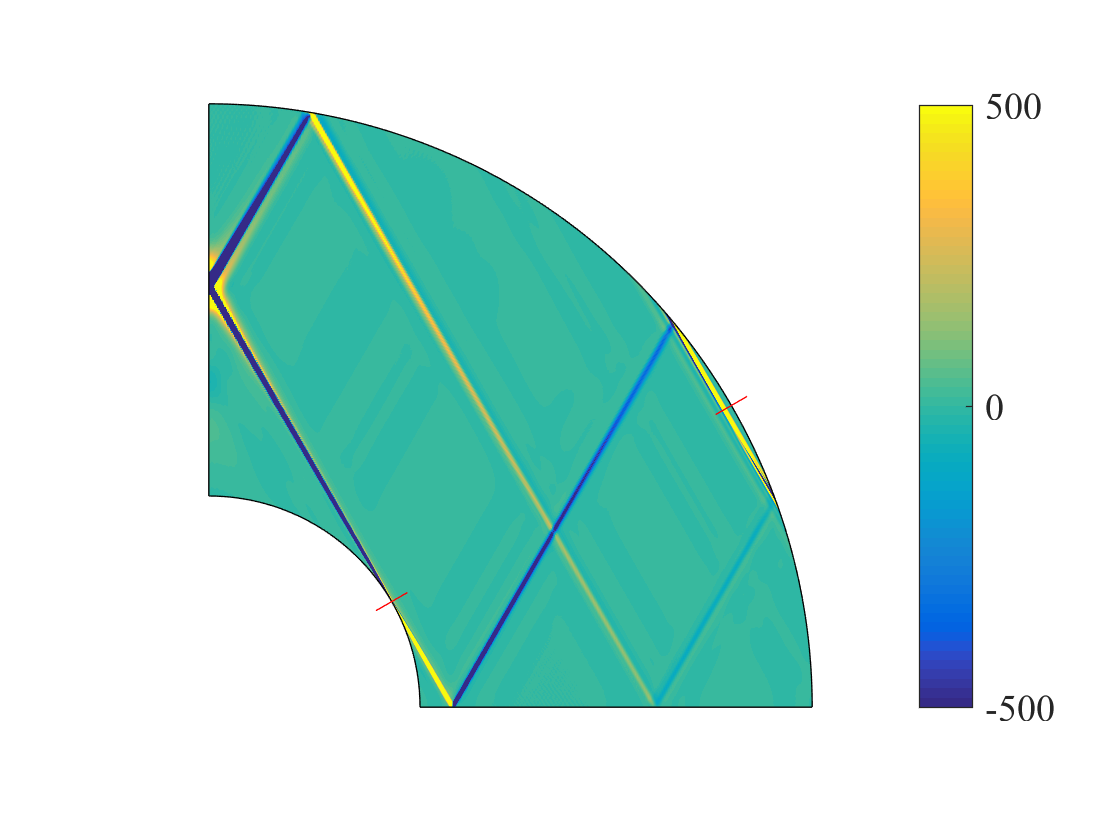}
\includegraphics[width=0.33 \textwidth,clip,trim=2cm 1.5cm 0.65cm 0cm]{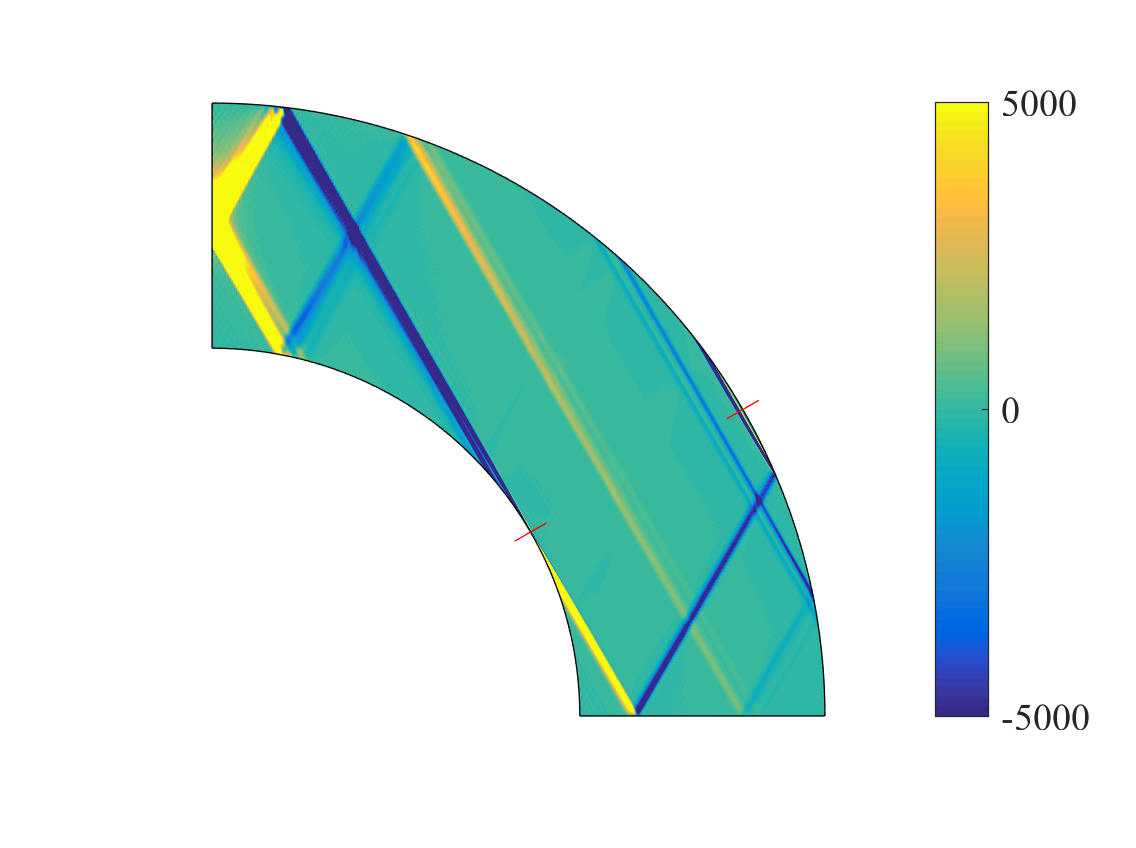}
\includegraphics[width=0.33 \textwidth,clip,trim=2cm 1.5cm 0.8cm 0cm]{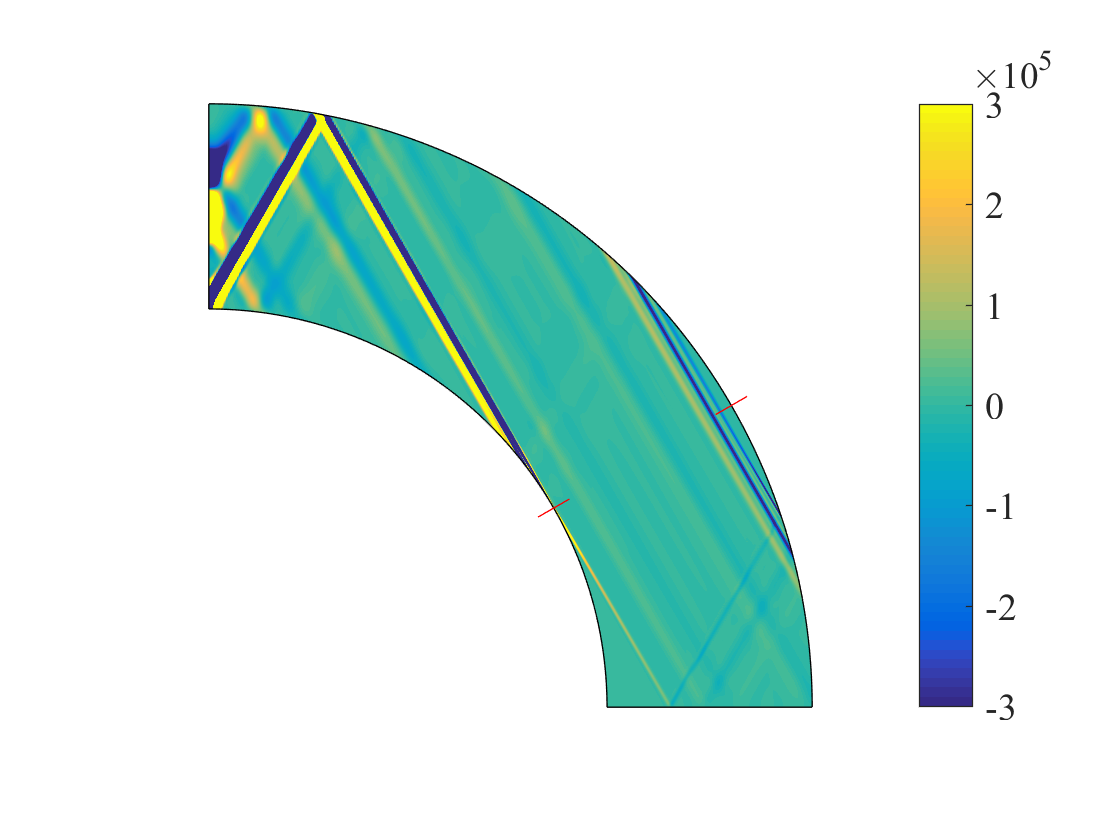} \\
(d) \hfill (e) \hfill(f) \\
\includegraphics[width=0.33 \textwidth,clip,trim=3cm 0cm 0cm 0cm]{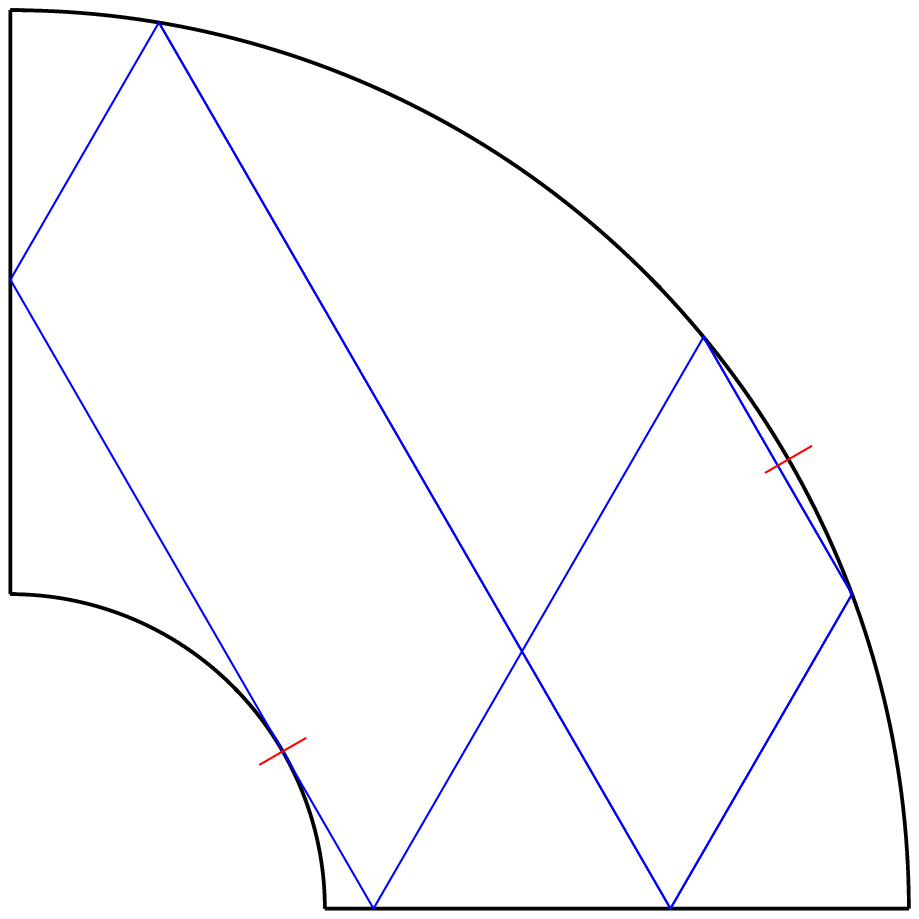}
\includegraphics[width=0.33 \textwidth,clip,trim=3cm 0cm 0cm 0cm]{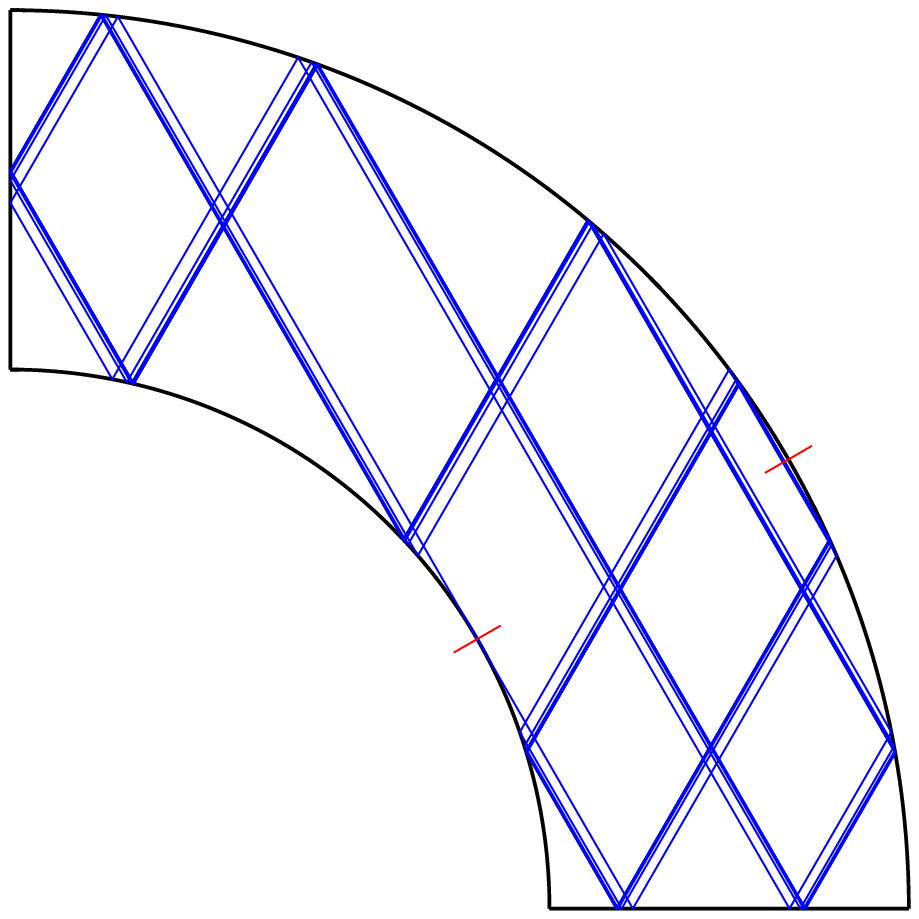}
\includegraphics[width=0.33 \textwidth,clip,trim=3cm 0cm 0cm 0cm]{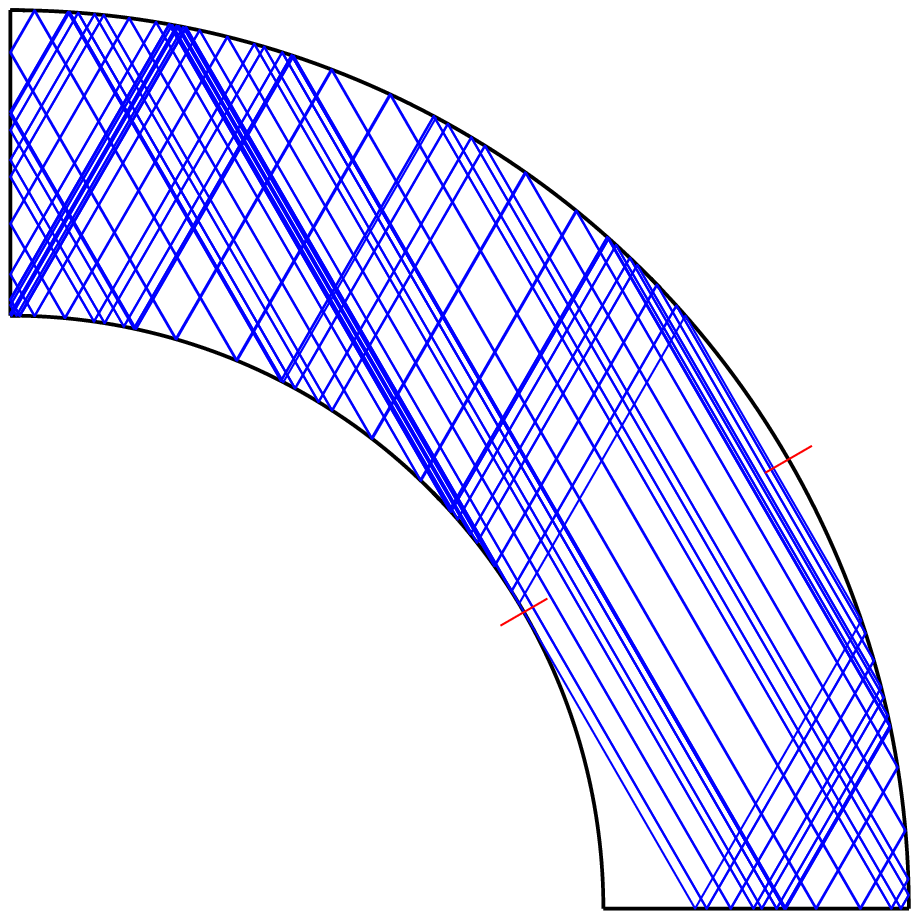} \\
(g) \hfill (h) \hfill(i) \\
\caption{Structure of the wavelike responses in the meridional plane for a homogeneous fluid. (a-c) Velocity $|\bmath{u}|$, (d-f) helicity $H=\bmath u \cdot \nabla \times \bmath u$, and (g-i) characteristics starting from the critical latitude at the inner core boundary. Red ticks indicate the critical latitudes. From left to right, the radius ratio $\eta=0.35$, 0.6, 0.66 respectively. $E=10^{-8}$. $L=800$ and $N=400$.}
\label{fig:ray_new}
\end{figure*}

\begin{figure}
\begin{center}
\includegraphics[width=0.45 \columnwidth,clip,trim=3cm 1cm 2cm 0.5cm]{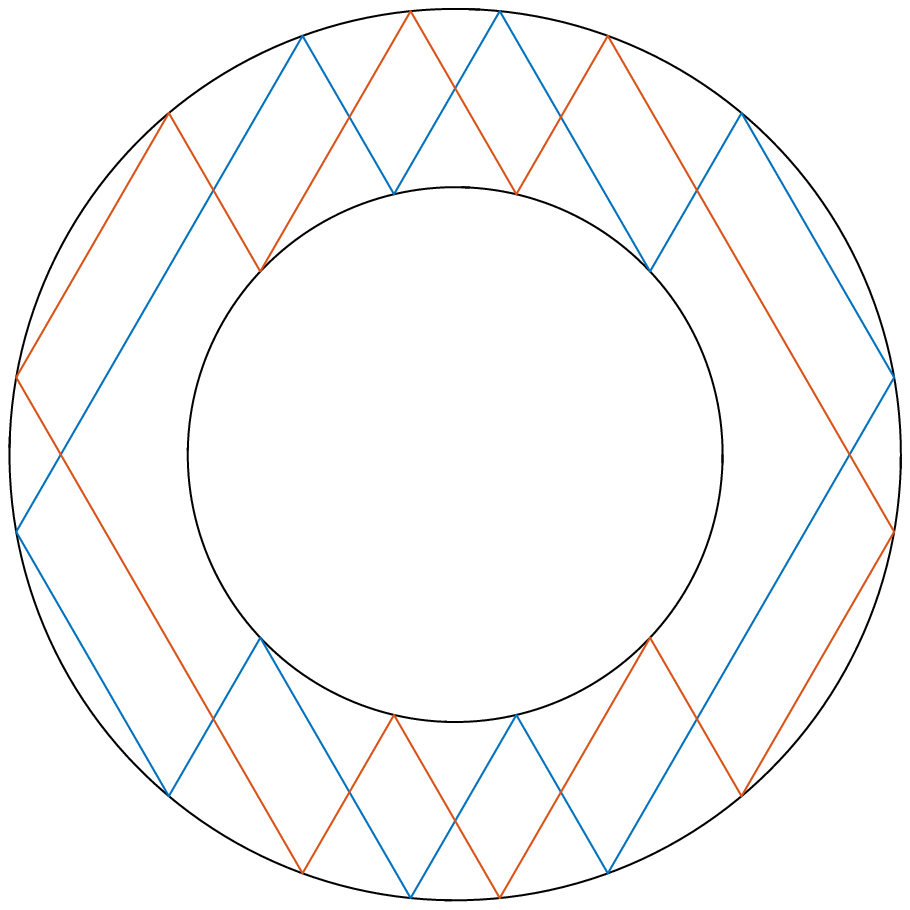}
\includegraphics[width=0.45 \columnwidth,clip,trim=3cm 1cm 2cm 0.5cm]{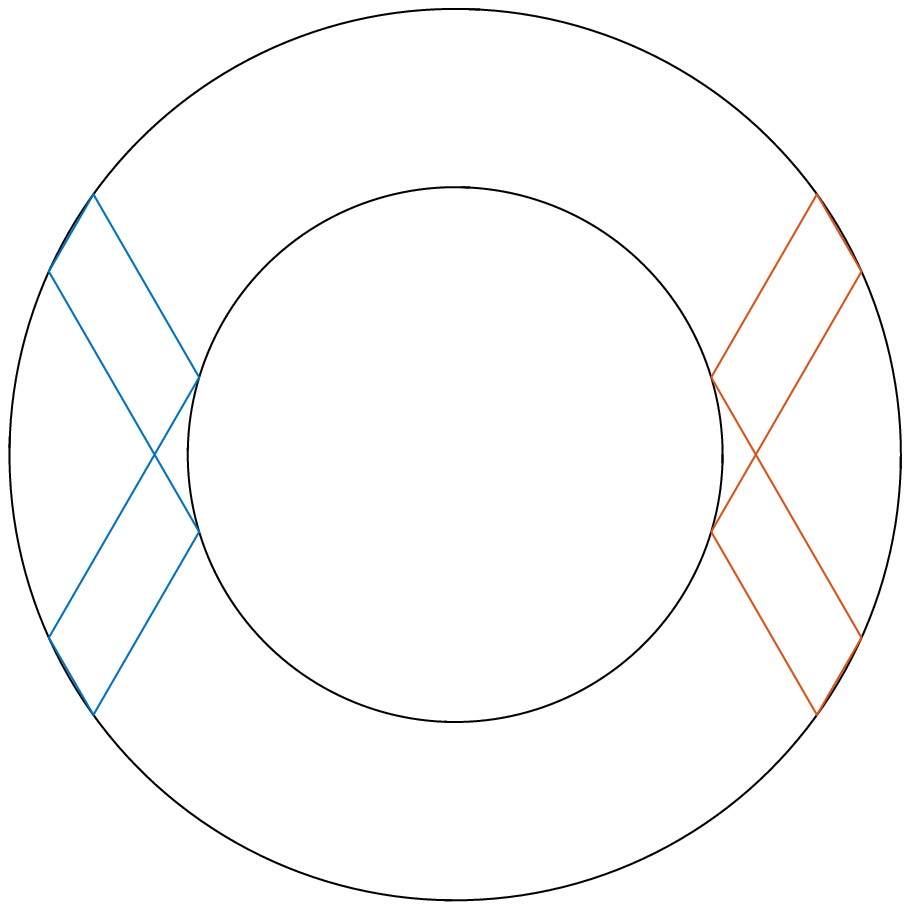} \\
(a) \hfill (b)
\end{center}
\caption{Inertial wave attractors in a spherical shell with $\eta=0.6$.}
\label{fig:Attractors}
\end{figure} 

Although the above equations are formulated for general barotropic fluids, they are also valid for a homogeneous fluid, which corresponds to the limit of the polytrope of index 0. We consider body 1 consisting of an incompressible homogeneous fluid and possibly containing a solid core of the same density as the fluid. In this case, we can find an analytic solution for the non-wavelike part \citep{Ogilvie2013MNRAS}
\begin{equation}
X=C \left[\left(\frac{r}{R}\right)^2 +\frac{2 \eta^5}{3}\left(\frac{R}{r}\right)^3 \right] Y_2^1(\theta,\phi) \mathrm{e}^{\mathrm{i} \Omega_s t},
\end{equation}
\begin{equation}
\bmath u_{nw}=-\nabla \dot{X}, 
\end{equation}
where 
\begin{equation}
 C=\frac{5}{4}\frac{A_{210} R^2}{(1-\eta^5)\omega_d^2}.
\end{equation}
The solvability condition determines the precession frequency 
\begin{equation}
 \Omega_p=-\frac{15}{8}\sqrt{\frac{5}{6 \pi}}\frac{A_{210}\Omega_s}{(1-\eta^5)\omega_d^2 \sin \alpha_s}.
\end{equation}
The negative precession frequency implies that the precession is retrograde as expected. The precession frequency determined by the solvability condition is consistent with the scaling $\Omega_p=O(\beta \epsilon)$, which we have used for the asymptotic analysis in section \ref{sec:equation}.

For the case of a whole homogeneous fluid body $\eta=0$, the precession frequency reduces to  
\begin{equation}
 \Omega_p=-\frac{15}{8}\frac{GM_2}{a^3}\frac{\Omega_s}{\omega_d^2}\cos i,
\end{equation}
if we assume that the spin angular momentum is much smaller than the orbital angular momentum ($\alpha_s=i$). This is equivalent to the conventional precession frequency given by equation (\ref{eq:omgp}) using $k_2=\frac{3}{2}$ and $k_*=\frac{2}{5}$ for a homogeneous body.

Substituting the precession frequency and the non-wavelike velocity into $\bmath{f_p}$ and $\bmath{f_{nw}}$, we find that the total force is curl-free for a whole fluid body ($\eta=0$):
\begin{equation}
\nabla \times (\bmath f_{nw}+\bmath f_{p})=0.
\end{equation}
This suggests that the wavelike response to the obliquity tide $\Psi_{210}$ is just a pressure perturbation due to precession with no internal waves being excited for a whole homogeneous fluid body. 
Therefore, there is no enhanced dissipation due to inertial waves.
 
However, the total force is not curl-free in the presence of a rigid inner core ($0<\eta<1$), i.e. $\nabla \times (\bmath f_{nw}+\bmath f_{p})\neq 0$. Therefore, we expect that non-trivial inertial waves can be excited by the precessional forcing and the Coriolis force due to the non-wavelike velocity.    

Fig. \ref{fig:ray_new} shows the spatial structures of the wavelike response in the meridional plane for different radii of the rigid core at the Ekman number $E=10^{-8}$. The top panel shows the velocity magnitude $|\bmath u|$, the middle panel shows the helicity $H=\bmath u \cdot (\nabla \times \bmath u)$ and the bottom panel shows characteristics of inertial waves starting from the critical latitude at the inner core boundary. Because of the peculiar dispersion relation of inertial waves, the characteristics are inclined by a fixed angle of $\arcsin (|\hat{\omega}/(2 \Omega_s)|)$ with respect to the rotation axis. At the critical latitude, which is also $\arcsin (|\hat{\omega}/(2 \Omega_s)|)$,  the characteristics are either tangential or perpendicular to the boundary. The obliquity tide has frequency $\hat{\omega}=-\Omega_s$ in the fluid frame, so the angle between the rotation axis and the characteristics is $30^{\circ}$ in all cases. In addition, inertial waves are helical and negative (positive) $H$ indicates inertial waves propagating upwards (downwards) \citep{Davidson2014GeoJI}. We can see that inertial waves emanate from the critical latitude at the inner core boundary and travel along the characteristics. However, the geometry of rays varies depending on the radius ratio $\eta$. 

For the case of $\eta=0.35$, the ray starting from the inner critical latitude forms a very simple orbit after a few reflections. Note that this type of ray path is neutral since the so-called Lyapunov exponent is zero \citep{Rieutord2001}. In other words, this is not a wave attractor although a closed ray path is formed after a few reflections. Such a simple ray beam is always excited so long as the radius ratio $\eta<0.5$ at the tidal frequency $|\hat{\omega}|/\Omega_s=1$.  

For $\eta=0.6$, there exist simple wave attractors as shown in Fig. \ref{fig:Attractors}. The helicity in Fig. \ref{fig:ray_new}(e) suggests that inertial wave beams are spawned from the inner critical latitude. The upward propagating beam (negative $H$)  emanating from the critical latitude converges to the inertial wave attractor in  Fig. \ref{fig:Attractors} (a) after multiple reflections, while the downward propagating beam (positive $H$)  converges to the inertial wave attractor in  Fig. \ref{fig:Attractors} (b).

For $\eta=0.66$, there is no simple wave attractor and the ray dynamics is rather complicated. However, the energy mainly concentrates around two characteristics between the north pole on the inner boundary and the inner critical latitude, where ray paths also show concentration. The helicity around the two characteristics has both negative and positive values, suggesting that wave beams bounce back and forth along these characteristics.

\begin{figure}
\begin{center}
\includegraphics[width=0.49 \textwidth]{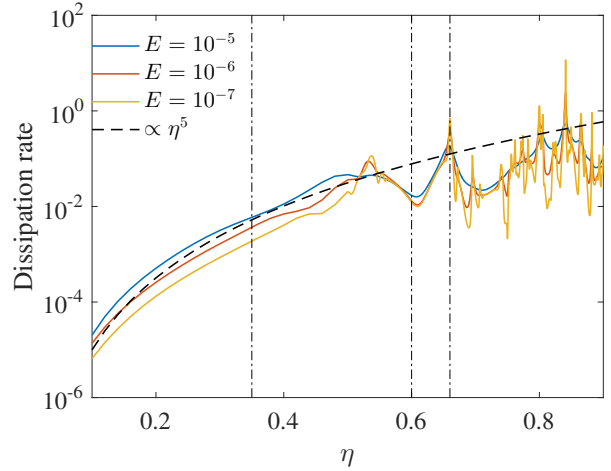} \\
\end{center}
\caption{Dimensionless dissipation rate $\tilde{D}$ versus radius ratio $\eta$ at $E=10^{-5}$, $E=10^{-6}$,$E=10^{-7}$ for a homogeneous fluid. Vertical lines indicate the values of $\eta$ shown in Fig. \ref{fig:ray_new}}
\label{fig:diss_as_ri_Ekman}
\end{figure}

Figure \ref{fig:diss_as_ri_Ekman} shows the dissipation rate as a function of the radius ratio at different Ekman numbers. For convenience, we show the dimensionless dissipation,
\begin{equation}\label{eq:Diss_N}
\tilde{D}=\frac{D}{\rho_0 \Omega_s^3 R^5 (A_{210}/\omega_d^2)^2}.
\end{equation}
which has been normalized by the tidal amplitude.  

The dissipation rate strongly depends on the radius ratio and roughly scales as $\eta^5$ (Fig. \ref{fig:diss_as_ri_Ekman}), which is in agreement with the core size dependence of the frequency averaged dissipation rate due to inertial waves \citep{Goodman2009ApJ,Ogilvie2009MNRAS,Ogilvie2013MNRAS,Rieutord2010}. When $\eta<0.5$, the ray dynamics exhibits similar behaviours as shown in Fig. \ref{fig:ray_new} (g), and the dissipation rate follows the scaling $\eta^5$ . However, when $\eta>0.5$, the dependence on the core size is more complicated, showing peaks and troughs. This can be attributed to different ray dynamics when  $\eta>0.5$. For example, the existence of wave attractors at $\eta=0.6$ leads to a trough in the dissipation rate. The case of $\eta=0.66$, in which wave beams concentrate into a certain region but do not form wave attractors, corresponds a peak in \ref{fig:diss_as_ri_Ekman}. In addition, we can see that the dissipation rate exhibits different Ekman-number dependence for different radius ratio.  

\begin{figure}
\begin{center}
\includegraphics[width=0.49 \textwidth]{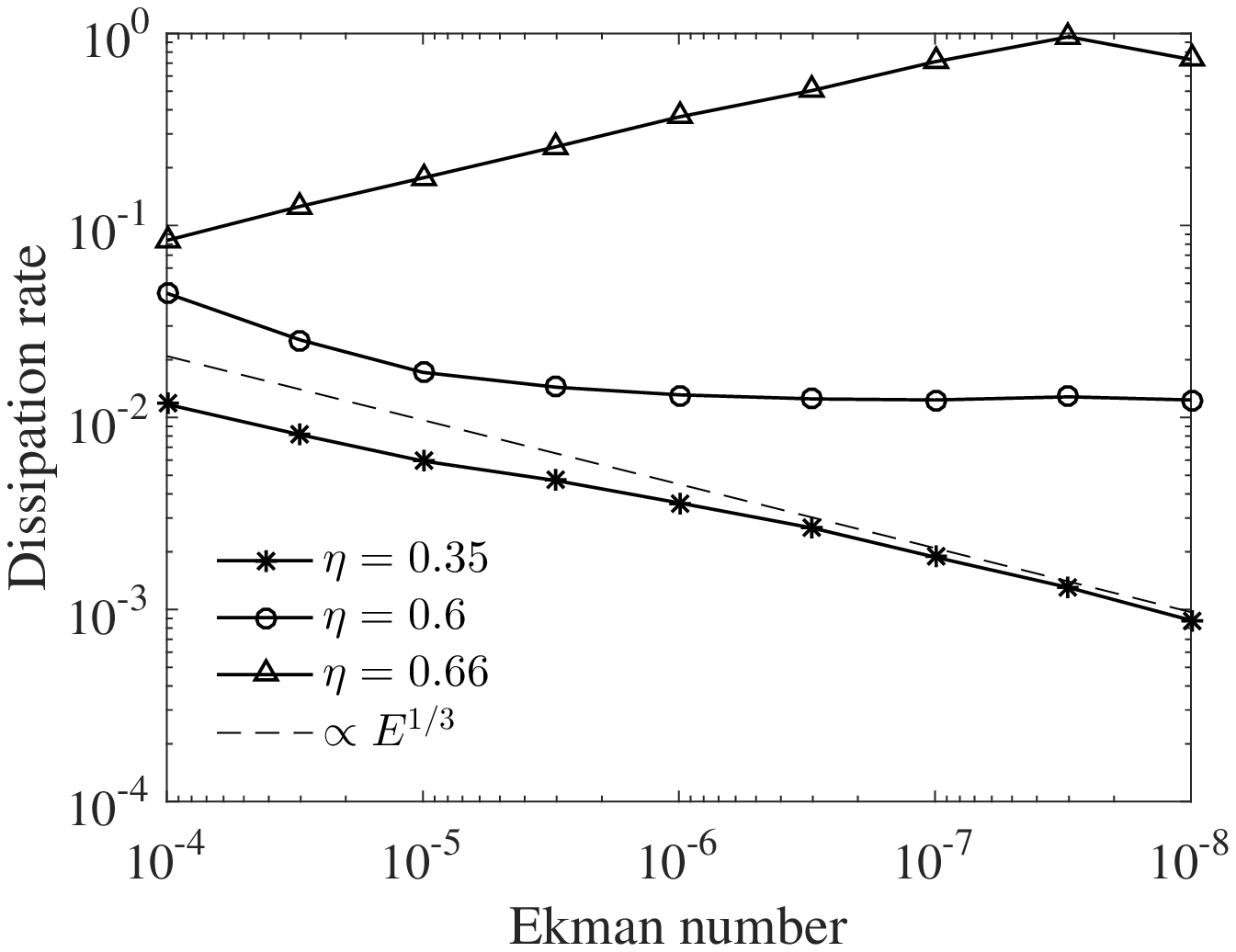}\\
(a)\\
\includegraphics[width=0.49 \textwidth]{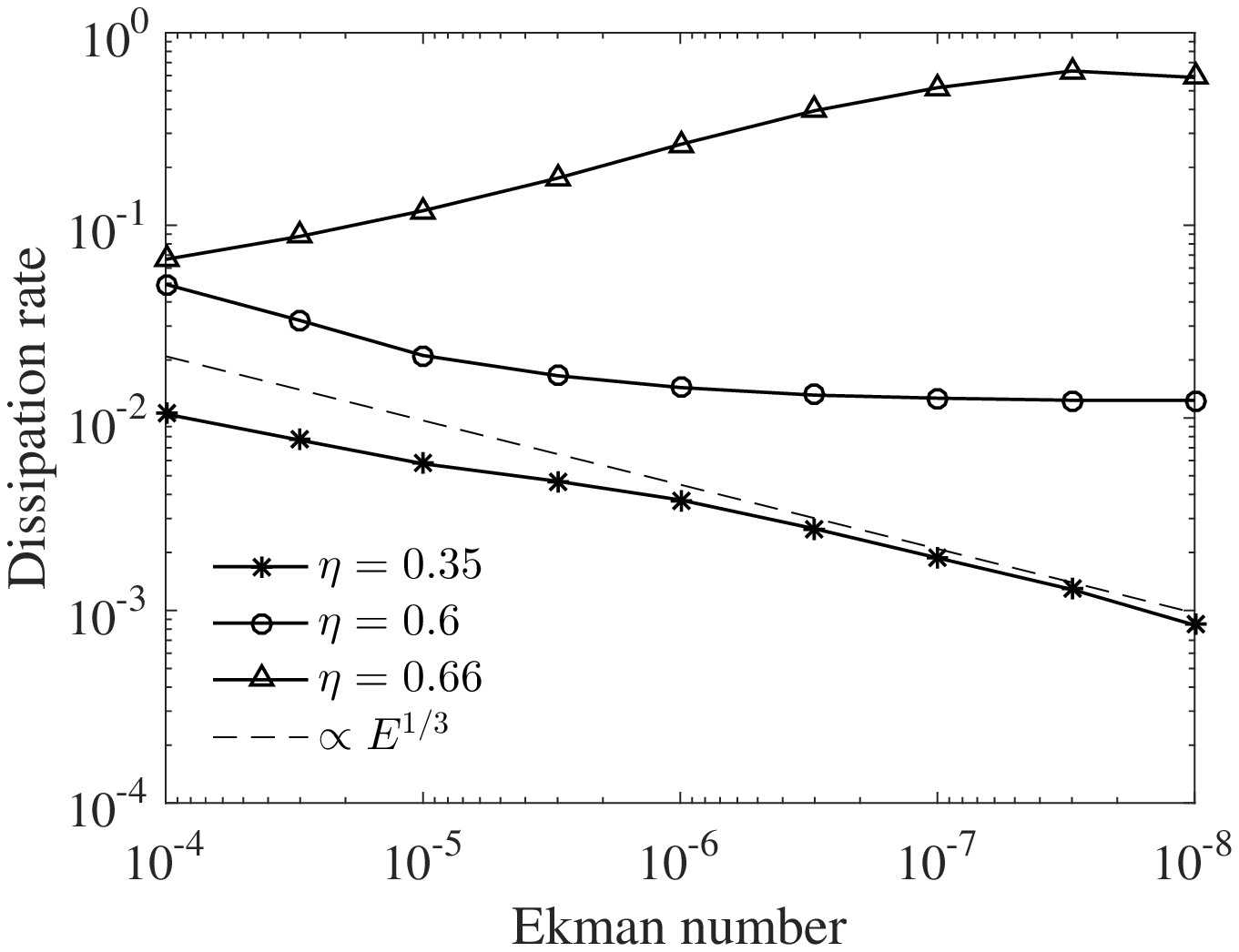}\\
(b)
\end{center}
\caption{Dimensionless dissipation rate $\tilde{D}$  versus Ekman number $E$ for a homogeneous fluid. (a) Stress free boundary condition and (b) no-slip boundary condition on the inner boundary.}
\label{fig:diss_as_Ekman_new}
\end{figure}

Fig. \ref{fig:diss_as_Ekman_new} shows the dissipation rate versus the Ekman number for three cases shown in Fig. \ref{fig:ray_new}, which represent three kinds of typical ray dynamics. The dissipation rate tends to scale as $E^{1/3}$ for the case $\eta=0.35$, where inertial waves beams emerging from the inner critical latitude form a simple closed ray path. The dissipation rate becomes independent of the Ekman number as we decrease $E$ for the case of $\eta=0.6$, where wave attractors are excited. This dependence is consistent with the scaling found by \citet{Ogilvie2005JFM} for wave attractors. At $\eta=0.66$, the dissipation rate increases on decreasing the Ekman number but appears to be saturated at very low Ekman numbers. In the case, the energy concentrates into a certain region and wave beams bounce back and forth without simple attractors. These results are in line with the numerical results of \citet{Ogilvie2009MNRAS} and \citet{Rieutord2010}, in which inertial waves are forced by other tidal components.      

Fig. \ref{fig:diss_as_Ekman_new} also compares the dissipation rate between the no-slip and stress free boundary conditions on the inner core. We can see very similar dissipation behaviours for the two different boundary conditions. The dissipation in the thin viscous boundary layer due to the no-slip boundary condition scales as $E^{1/2}$, which is negligible compared to the dissipation due to wave beams in the bulk. This suggests that the dissipation in our models is mainly contributed by the localized inertial waves in the bulk. 

Similar inertial waves also emerge from the critical latitudes in precession flows confined in a rigid container \citep{Hollerbach1995,Tilgner1999}, like the liquid core confined in the precessing mantle of the Earth. In such case, inertial waves are spawned because of the singularity of the Ekman pumping at the critical latitudes \citep{Kerswell1995}. The dissipation of inertial waves in the bulk is negligible compared to the dissipation in the Ekman boundary layer\citep{Hollerbach1995}. In the model we consider here, inertial waves are forced by the body force $\bmath{f_p}+\bmath{f_{nw}}$, although wave beams seem to emanate from the critical latitudes as well. The dissipation mainly arises from localized inertial waves in the bulk. The Coriolis force $\bmath{f_{nw}}$ due to the non-wavelike motion contains $l=1$, $l=2$ and $l=3$ components (see equations \ref{eq:proj1}-\ref{eq:proj3}), since $\bmath{u_{nw}} \propto Y_2^1(\theta, \phi)$. The $l=1$ component in $\bmath{f_{nw}}$ provides the torque for the precessional motion (or the spin-over mode), whereas non-trivial inertial waves are forced by $l=2$ and $l=3$ components in $\bmath{f_{nw}}$ effectively.  

\subsection{In a polytrope of index 1}
We now consider a fluid polytrope of index 1. There may exist a rigid core, which has the same polytropic density profile as the whole body is fluid. The density profile is given as \citep{Ogilvie2004ApJ}
\begin{equation}
\rho_0=\left( \frac{\pi M_1}{4 R^3}\right) \frac{\sin(kr)}{kr},
\end{equation}
where $k=\pi/R$. 
The gravitational acceleration is
\begin{equation}
g=\frac{G M_1}{\pi r^2}[\sin(kr)-kr\cos(kr)].
\end{equation}
The sound speed can be obtained from
\begin{equation}
\frac{\rho_0}{V_s^2}=-\frac{1}{g} \frac{d \rho_0}{d r}=\frac{\pi}{4 G R^2}.
\end{equation}

The non-wavelike part can be obtained by numerically solving the ODEs (\ref{eq:nw_Phi_ODE}-\ref{eq:nw_X_ODE}). Then we can determine the precession frequency using the solvability condition. For a whole polytropic fluid, we find the precession frequency
\begin{equation}
 \Omega_p=-\frac{3(15-\pi^2)}{4(\pi^2-6)}\frac{GM_2}{a^3}\frac{\Omega_s \cos i}{\omega_d^2},
\end{equation}
provided that the spin angular momentum is much smaller than the orbital angular momentum ($\alpha_s=i$). This is also consistent with the precession frequency given by equation (\ref{eq:omgp}), considering $k_2=\frac{15-\pi^2}{\pi^2}$  and $k_*=\frac{2(\pi^2-6)}{3 \pi^2}$ for a polytrope of index $n=1$ \citep{Brooker1955MNRAS}.

\begin{figure*}
\includegraphics[width=0.33 \textwidth,clip,trim=2cm 1.5cm 1cm 0cm]{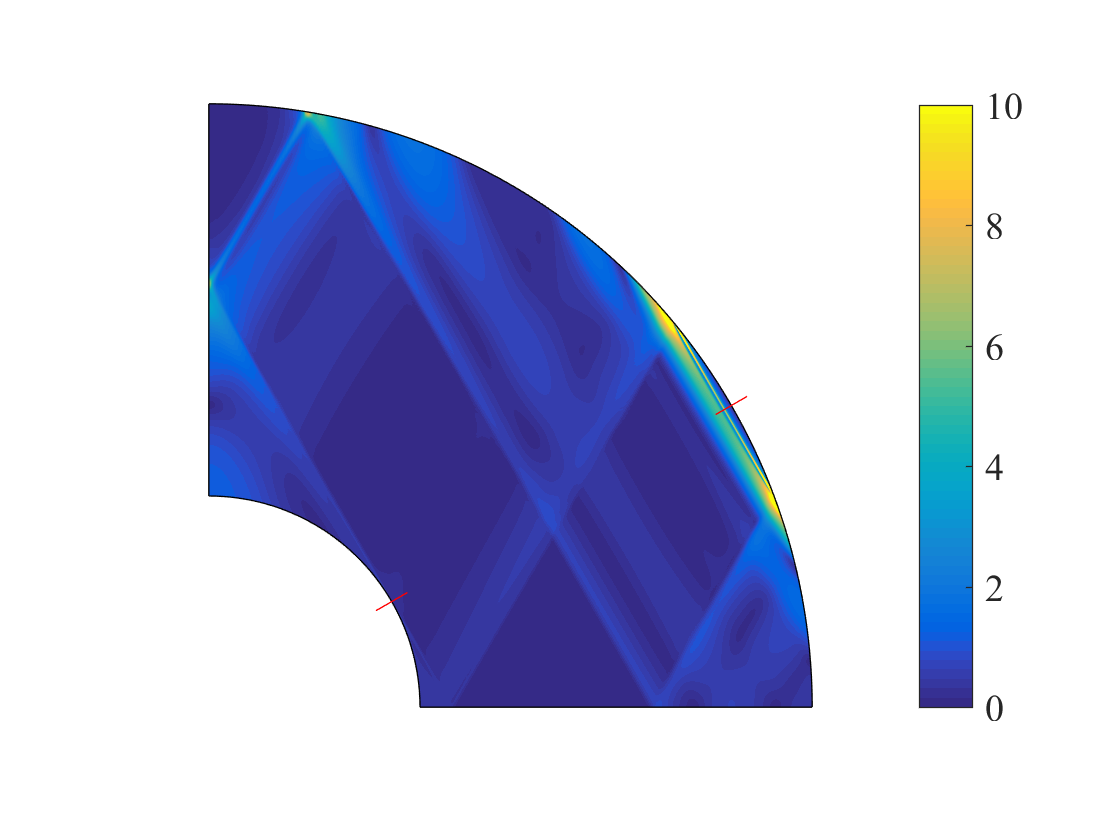}
\includegraphics[width=0.33 \textwidth,clip,trim=2cm 1.5cm 1cm 0cm]{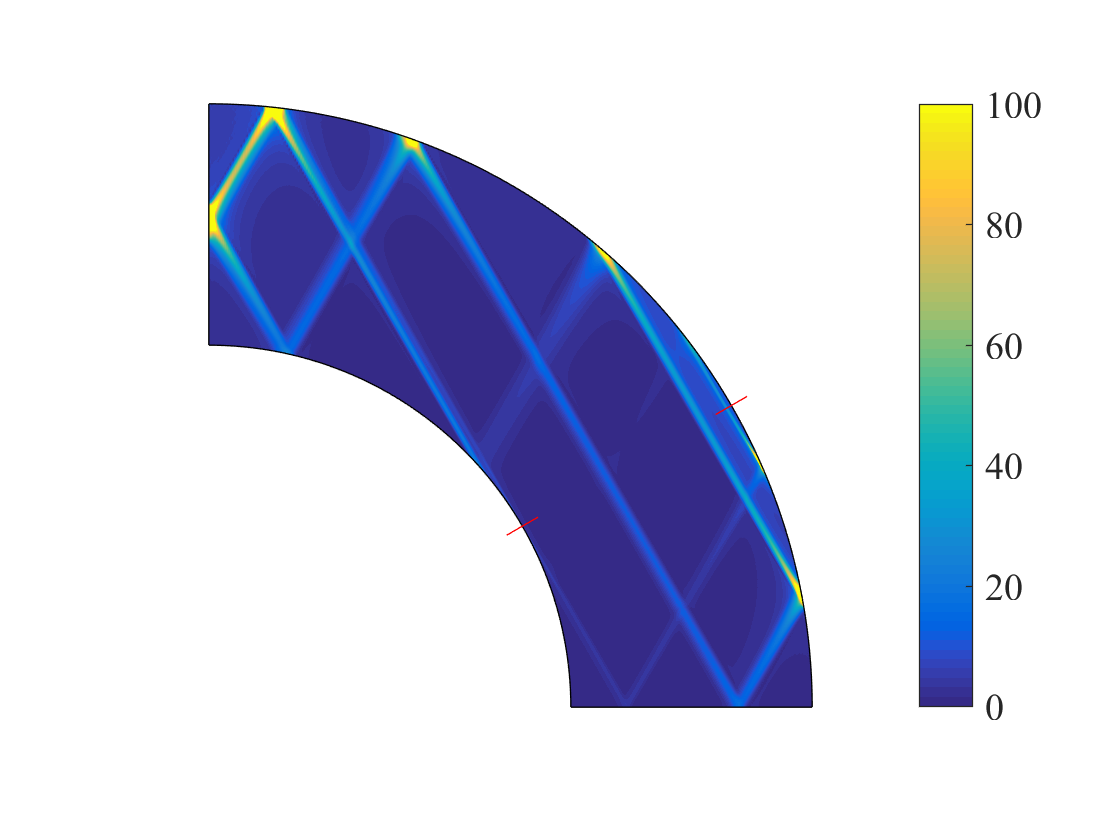}
\includegraphics[width=0.33 \textwidth,clip,trim=2cm 1.5cm 1cm 0cm]{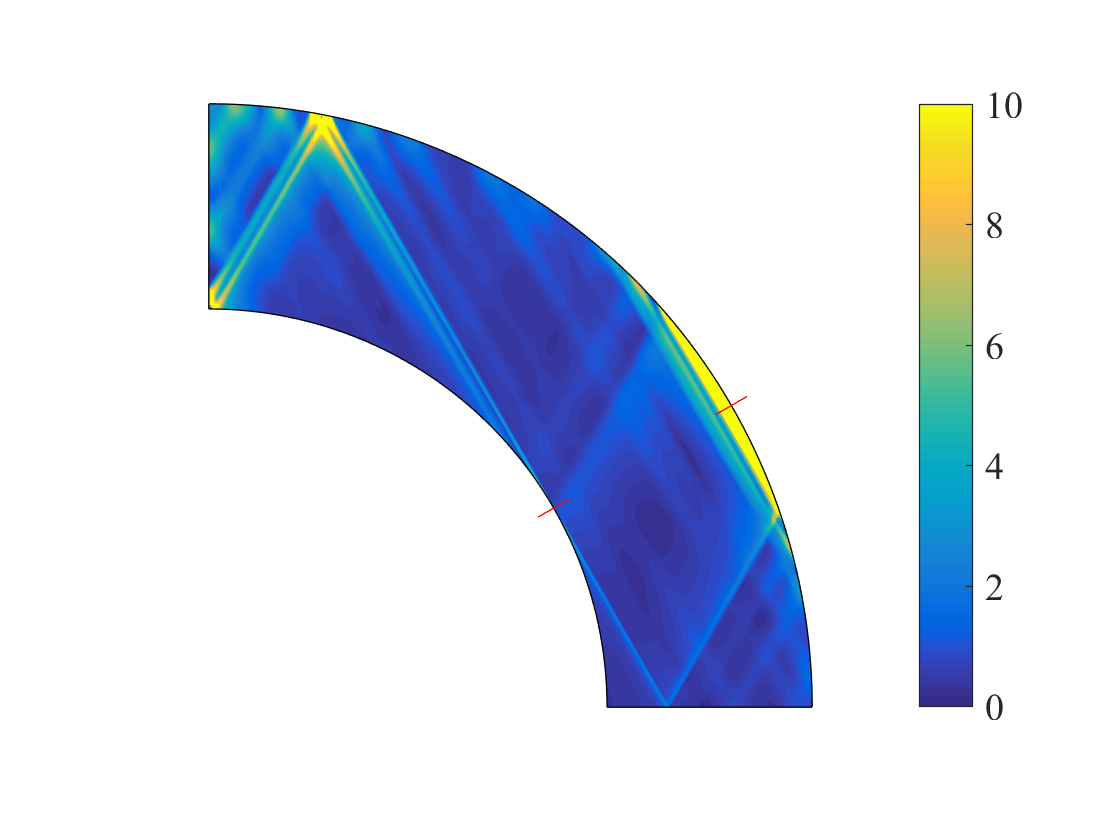} \\
(a) \hfill (b) \hfill(c) \\
\caption{Structure of the wavelike velocity $|\bmath{u}|$ in the meridional plane for a fluid polytrope of index 1.  (a)$\eta=0.35$, (b)$\eta=0.6$, (c)$\eta=0.66$. $E=10^{-8}$. $L=800$ and $N=400$.}
\label{fig:ur_poly}
\end{figure*}
\begin{figure}
\begin{center}
\includegraphics[width=0.49 \textwidth]{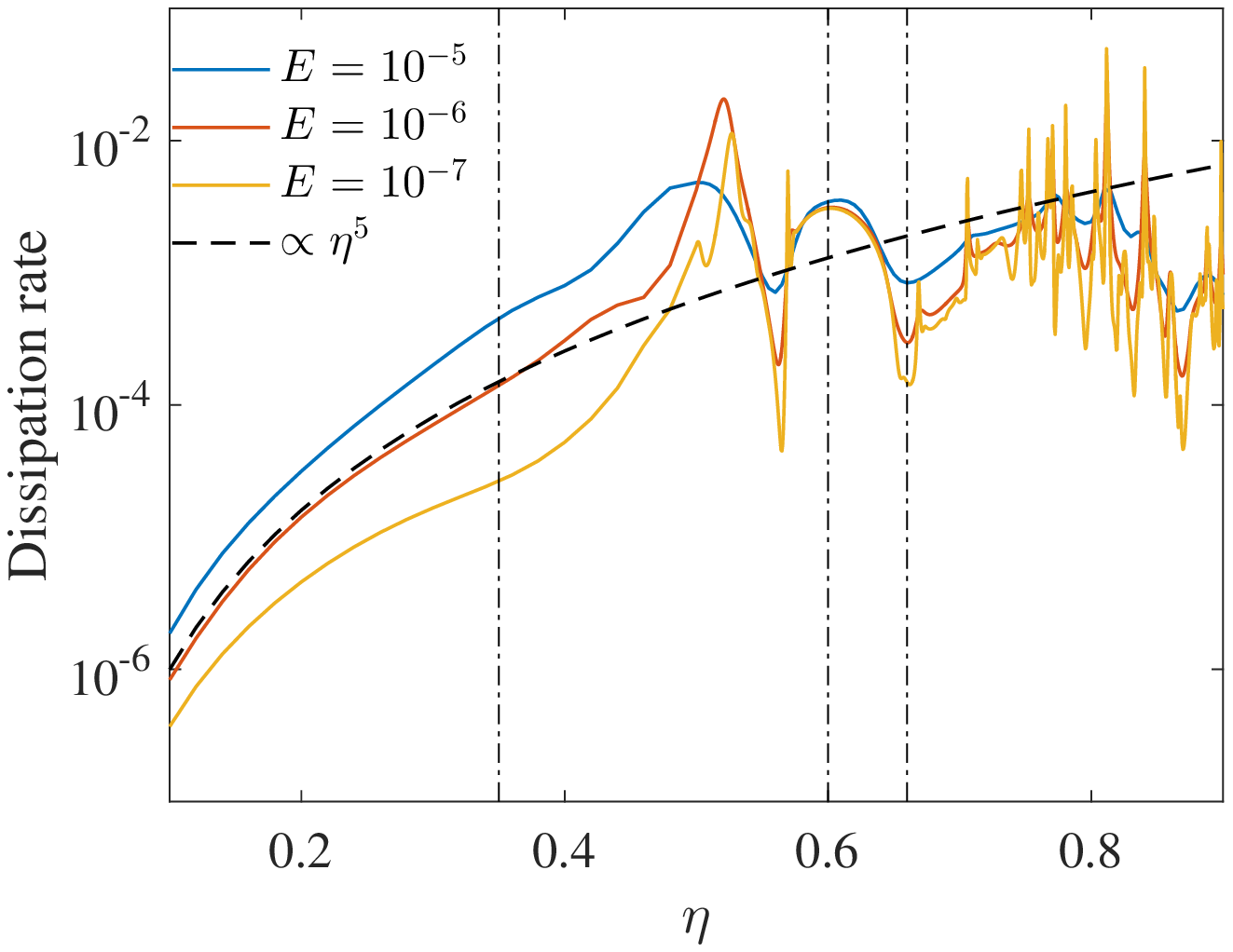}\\
(a)\\
\includegraphics[width=0.49 \textwidth]{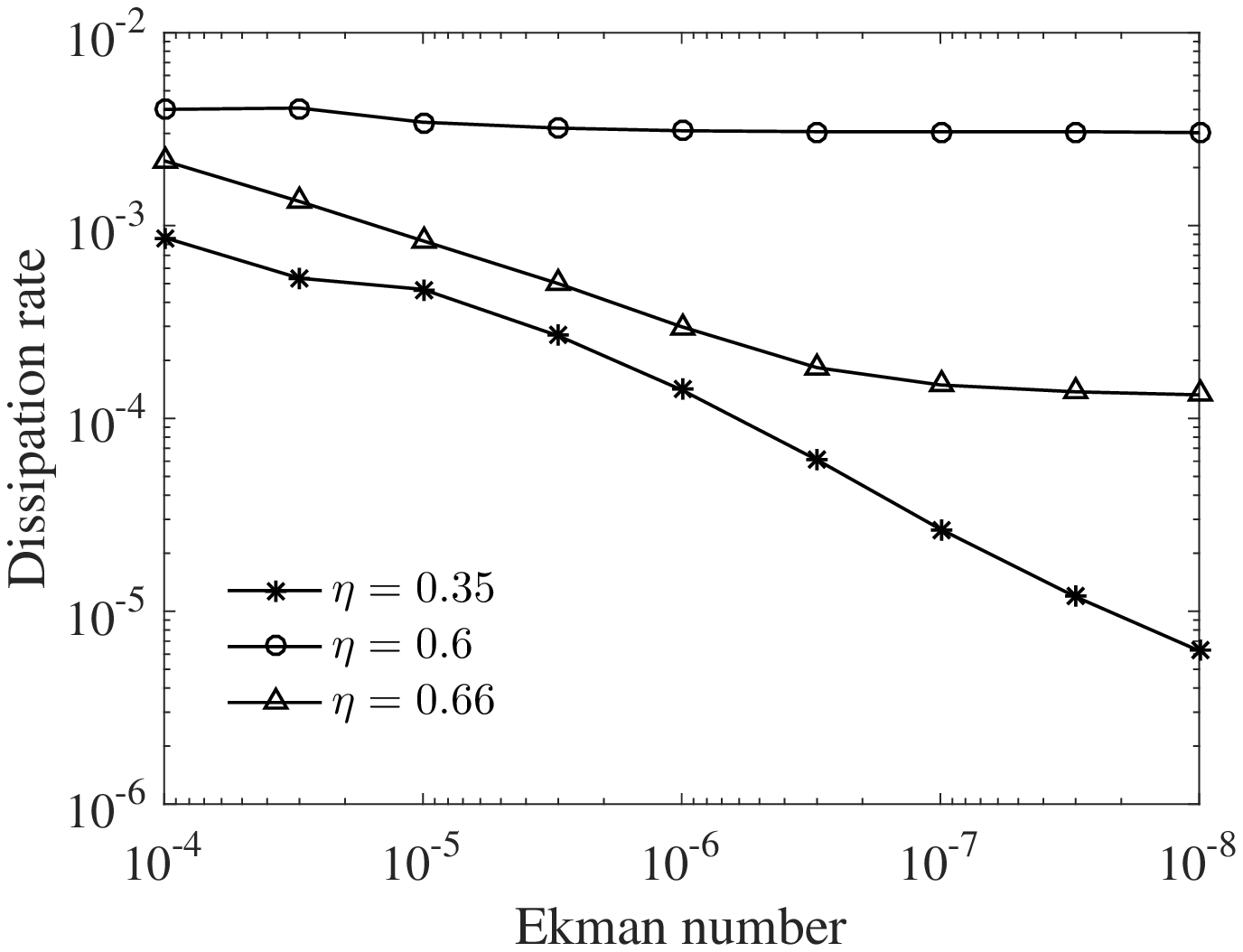}\\
(b) 
\end{center}
\caption{Dissipation rate versus (a) radius ratio and (b) Ekman number for a fluid  polytrope of index 1.}
\label{fig:Diss_poly}
\end{figure}

Once the precession frequency and the non-wavelike part are determined, the wavelike part can be solved as we described above. Fig. \ref{fig:ur_poly} shows the wavelike velocity structure in the meridional plane for the radius ratio $\eta=0.35$, $\eta=0.6$ and $\eta=0.66$.  We can see that the velocity structures are very similar to that of a homogeneous fluid in Fig. \ref{fig:ray_new}. However, for the case of $\eta=0.35$, an additional wave beam appears, which is weaker and thicker and  is associated with the north pole on the inner core boundary.

Fig. \ref{fig:Diss_poly} shows the dimensionless dissipation rate as a function of the radius ratio (a) and the Ekman number (b). In the polytropic fluid, the dimensionless dissipation rate is given as
\begin{equation}\label{eq:Diss_N}
\tilde{D}=\frac{D}{\bar{\rho} \Omega_s^3 R^5 (A_{210}/\omega_d^2)^2},
\end{equation}
where $\bar{\rho}=3M_1/(4\pi R^3)$ is the mean density. 

Fig. \ref{fig:Diss_poly}(a) suggests that the dissipation rate is proportional to $\eta^5$ when $\eta<0.5$, just as in the case of the homogeneous fluid. However, the peaks and troughs in the range $\eta>0.5$ seem to be opposite of those  in the case of the homogeneous fluid. For instance, the dissipation rate exhibits a peak around $\eta=0.6$ for the case of the polytrope of index 1 whereas it corresponds a trough for the case of the homogeneous fluid. Correspondingly, there is a peak at $\eta=0.66$ in the homogeneous fluid but it is a trough in the polytrope of index 1. 
This suggests that the shape of the dissipation curves cannot be explained by the ray dynamics alone, which has been noted by \cite{Ogilvie2009MNRAS}.    

Fig. \ref{fig:Diss_poly}(b) shows the dependence of the dissipation rate on the Ekman number at $\eta=0.35$, $\eta=0.6$ and $\eta=0.66$. The dissipation rate at $\eta=0.35$ decreases as the Ekman number is  decreased, which is similar to that of the homogeneous fluid. However, the strongest dissipation occurs at $\eta=0.6$, where inertial wave attractors are excited. The dissipation rate at $\eta=0.66$ decreases on reducing the Ekman number, but seems to converge to a low level at low Ekman numbers ($E<10^{-7}$).

In summary, our numerical calculations show that dissipative inertial waves can be excited by the obliquity tide if there exists a rigid core that excludes the waves. The velocity structures of inertial waves are similar between the homogeneous fluid and the fluid polytrope of index 1. However, the dissipation rate resulting from inertial waves depends on the density profile, the core size and also the Ekman number for some cases. We will roughly estimate the tidal quality factor of the obliquity tide based on these calculations and discuss some astrophysical implications in the next section.

\section{Discussion} \label{sec:Diss}
\begin{figure}
\includegraphics[width=0.5\textwidth]{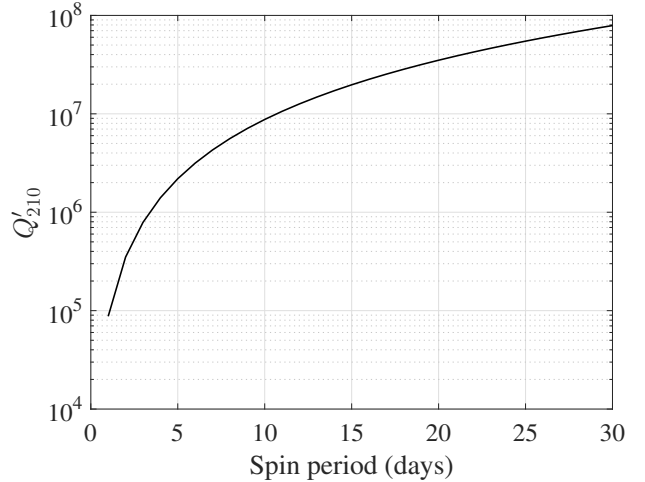}
\caption{Tidal quality factor $Q'_{210}$ as a function of the spin period for a solar-type star. The calculation is based on equation (\ref{eq:Q210}) and take $\tilde{D}=10^{-3}$.}
\label{fig:Q210}
\end{figure}
The tidal dissipation is conventionally parameterized as the tidal quality factor, which depends on the tidal frequency and amplitude in general. For a tidal component $\Psi_{lmn}=A_{lmn}r^l Y_l^m(\theta,\phi)\mathrm{e}^{\mathrm{i}\hat{\omega}t}$, the dissipation rate due to this can be related to the tidal quality factor as \citep{Ogilvie2014}
\begin{equation}
 D=|\hat{\omega}| \frac{(2l+1)R^{2l+1}|A_{lmn}|^2}{8 \pi G} \frac{3}{2}\frac{1}{Q_{lmn}'},
\end{equation}
where $Q'_{lmn}=\frac{3}{2k_2}Q_{lmn}$ is the modified tidal quality factor associated with the tidal component $\Psi_{lmn}$.

For the obliquity tide $\Psi_{210}=A_{210}r^2 Y_l^m \mathrm{e} ^{\mathrm{i}\Omega_st}$, we obtain the tidal quality factor
\begin{equation}\label{eq:Q210}
\frac{1}{Q'_{210}}=\frac{4}{5}\frac{\Omega_s^2}{\omega_d^2} \tilde{D},
\end{equation}
where $\tilde{D}$ is the dimensionless dissipation rate defined in equation (\ref{eq:Diss_N}) and plotted in Figs. \ref{fig:diss_as_ri_Ekman}, \ref{fig:diss_as_Ekman_new} and \ref{fig:Diss_poly}. The tidal quality factor is independent of the tidal amplitude as we consider only linear tides here. The dissipation rate due to the excitation of inertial waves by the obliquity tides depends on the internal structure (e.g. core size, density profile), as well as the Ekman number. Nevertheless, taking a typical value of $\tilde{D}\approx 10^{-3}$ according to our calculations, Fig. \ref{fig:Q210} shows the tidal quality factor $Q'_{210}$ as a function of the spin period of a solar-type star. We can see that the tidal quality factor depends much on the stellar spin frequency as noted by \cite{Ogilvie2007ApJ} and \citet{Ogilvie2009MNRAS}. For a solar-type star with the spin period of 10 days, the tidal quality factor due to the obliquity tide $Q'_{210}\approx10^{7}$, which is about three orders of magnitude smaller than the tidal quality factor without the excitation of inertial waves, e.g. $Q'\approx 10^{10}$ \citep{Ogilvie2007ApJ}. Although solar-type stars do not have rigid cores, they do have stably-stratified radiative cores that present a nearly rigid boundary to inertial waves in the convective envelop \citep{Ogilvie2004ApJ}.

It has been suggested that tidal dissipation may have played an important role in reducing the stellar obliquities in hot Jupiter systems \citep{Winn2010ApJ}. If different tidal components involve a common tidal quality factor for the host star, the spin-orbit misalignment and the orbit would decay on a similar time scale since the spin angular momentum and the orbital angular momentum are comparable in hot Jupiter systems. The planet would be destroyed if the tidal dissipation is effective in damping the stellar obliquity. However, the tidal quality factor associated with the obliquity tide may be very different from that of the orbital evolution. 
 For typical parameters of hot Jupiter systems (e.g. the orbital period is $\sim$ 1 day and the stellar spin period is $\sim$ 10 days), the frequency of the obliquity tide is within the range of inertial waves whereas the frequencies of other tidal components are beyond the spectrum of inertial waves as noted by \citet{Lai2012MNRAS}. Therefore, the obliquity tide can achieve a much lower tidal quality factor (e.g. $Q'_{210}\approx10^{7}$)  than that of other tidal components (e.g. $Q'\approx 10^{10}$), which are unable to excite inertial waves.   
This suggests that the spin-orbit alignment can proceed much faster than the orbital decay in hot Jupiter systems because of the smaller tidal quality factor of the obliquity tide $Q'_{210}$, which hardly affects the orbital evolution. Indeed, some recent studies \citep{Rogers2013ApJ,Xue2014ApJ,Li2016ApJ} have assumed $Q'_{210}\ll Q'$ to investigate tidal interactions in hot Jupiter systems, showing that the stellar obliquity can evolve on a shorter time scale than the orbital decay. However, \citet{Lai2012MNRAS} and \cite{Rogers2013ApJ} have noted that the enhanced dissipation of the obliquity tide ($Q'_{210}\ll Q'$) may lead to the spin-obit angle stalling around $90^{\circ}$ or $180^{\circ}$ in many systems, which is incompatible with observations. \citet{Xue2014ApJ} and \citet{Li2016ApJ} suggested that the system can evolve out of the retrograde or polar orbits due to the dissipation of other tidal components and eventually approach the aligned configuration.    

There are no observational constraints on the obliquities of exoplanets so far. If the planetary spin axis is misaligned with respect to the orbital normal, similar inertial waves would be also excited by the obliquity tide in gaseous planets with a rocky core. In hot Jupiters, however, strong nonlinear effects such as precessional instabilities may be the dominant dissipation mechanism because of relatively rapid axial precession of the planets \citep{Barker2016MNRAS}.

There is a parallel analogy between the obliquity tide and the eccentricity tides in synchronized binary star systems, which have frequencies $\hat{\omega}=\pm \Omega_s$ in the fluid frame and can excite inertial waves \citep{Ogilvie2007ApJ}. The stellar tidal quality factor governing the binary circularization can be significantly smaller than the tidal quality factor of the host star governing the orbital decay in hot Jupiter systems for the same reason. This may explain the dichotomy of the stellar tidal quality factor for the hot Jupiter and the binary circularization problem as discussed by \citet{Ogilvie2007ApJ}. 

Although the obliquity tide and the eccentricity tides may lead to comparable tidal quality factors in synchronized binary stars, the spin-orbit alignment should proceed faster than the circularization since the orbital angular momentum is much larger the spin angular momentum in binary star systems. Recently, great efforts, including the BANANA project \citep{Albrecht2009Nature,Albrecht2014ApJ} and the EBLM project \citep{Triaud2013AA},  have been made to measure stellar obliquities in eclipsing binary star systems using the Rossiter-McLaughlin effect. Among a few systems with measurements of the spin-orbit angle (sky-projected), two systems have been found showing significant spin-orbit misalignments:  DI Herculis and CV Velorum. In DI Herculis, the spin axes of both stars are almost perpendicular to the orbital axis and the orbital eccentricity is 0.49 \citep{Albrecht2009Nature}. In CV Velorum, the primary and secondary stars have sky-projected obliquities of $-52^{\circ}$ and $3^{\circ}$ respectively and the orbit is circular \citep{Albrecht2014ApJ}. The tidal effect alone seems to have difficulty to reconcile the observed obliquities and eccentricities in these systems \citep{Albrecht2014ApJ}. Scenarios involving a third body have been considered recently to understand the eccentricity and the spin-orbit misalignment in binary star systems \citep{Anderson2016arXiv}.

\section{Conclusions}\label{sec:Conc}
Motivated by understanding the role of the obliquity tide in the evolution of the spin-orbit misalignment, we have studied the tidal interactions in spin-orbit misaligned systems. We formulated a set of linearized equations governing the tidal responses in barotropic fluid bodies by taking into account the mutual precession of the spin axis and orbital axis around the total angular momentum vector. The linearized equations are decomposed into the \textit{non-wavelike} and \textit{wavelike} parts and numerically solved using a pseudo-spectral method in spherical geometries. Numerical solutions in a homogeneous fluid and in a polytrope of index 1 have shown that non-trivial inertial waves can be excited on top of precession by the obliquity tide in the presence of a rigid core. Inertial waves are forced by the precessional forcing and the Coriolis force due to the non-wavelike motion and lead to enhanced tidal dissipation.  We estimated the tidal quality factor associated with the obliquity tide $Q'_{210}\approx 10^7$ for a solar-type star with spin period of 10 days, which is about three orders of magnitude smaller than those of other tidal components if their frequencies are outside the frequency range of inertial waves. For typical parameters of hot Jupiter systems, the frequency of the tidal forcing governing the orbital evolution is well beyond the spectrum of inertial waves. Therefore, it is possible that the damping of the spin-orbit misalignment can be much faster than the orbital decay due to the excitation of inertial waves by the obliquity tide as suggested by \citep{Lai2012MNRAS}. Our study has demonstrated the excitation of inertial waves by the obliquity tide and allows us to roughly estimate the corresponding tidal quality factor.   
  
We have adopted a simplified model in this study to demonstrate that non-trivial inertial waves can be excited by the obliquity tide on top of precession in spin-orbit misaligned systems. More realistic stellar models should be considered in future to account for the role of tidal dissipation in the distribution and evolution of the spin-orbit misalignments in exoplanetary systems and binary stars. 

\section*{Acknowledgements}

We would like to thank Adrian Barker for useful discussions. Y.L acknowledges the Swiss National Science Foundation for a PostDoc Mobility fellowship. This study is also supported by the  Isaac Newton Trust. 

%%%%%%%%%%%%%%%%%%%%%%%%%%%%%%%%%%%%%%%%%%%%%%%%%%



\bibliographystyle{mnras}
\bibliography{reference} % if your bibtex file is called example.bib

%%%%%%%%%%%%%%%%%%%%%%%%%%%%%%%%%%%%%%%%%%%%%%%%

%%%%%%%%%%%%%%%%% APPENDICES %%%%%%%%%%%%%%%%%%%%%

\appendix
\section{Projection of the equations onto Spherical Harmonics} \label{app:projection}
Substituting equations~ (\ref{eq:SphTor_ur}-\ref{eq:SphTor_W}) into the governing equations (\ref{eq:wave_NS}-\ref{eq:wave_mass}) for the wavelike part and projecting on to spherical harmonics, we obtain 
\begin{multline} \label{eq:proj1}
  \mathrm{i}\Omega_s a_n-2\mathrm{i}\Omega_s r b_n+2\Omega_s r[(n-1)q_nc_{n-1}-(n+2)q_{n+1}c_{n+1}] \\=-\frac{d W_n}{d r}
  -\frac{n(n+1) \nu}{r^2}\left[a_n-\frac{d}{d r}(r^2 b_n)\right]+2 \Omega_s^2 \frac{X(r)}{r}\delta_{2n},
\end{multline}
\begin{multline}\label{eq:proj2}
  \mathrm{i}\Omega_s r^2 b_n-\frac{2\mathrm{i}\Omega_s}{n(n+1)}\left(r a_n+r^2b_n\right)\\+2\Omega_s r^2\left[\frac{n-1}{n}q_nc_{n-1}+\frac{n+2}{n+1}q_{n+1}c_{n+1}\right] \\
  = -W_n+\nu\left[\frac{2a_n}{r}+\frac{1}{r^2}\frac{d}{d r}(r^4\frac{d b_n}{d r})-(n-1)(n+2)b_n\right] \\
  +\frac{2\Omega_s^2}{n(n+1)}\left(r \frac{\mathrm{d} X(r)}{\mathrm{d} r}+X(r)\right)\delta_{2n},
\end{multline}
\begin{multline} \label{eq:proj3}
  \mathrm{i}\Omega_s r^2c_n-\frac{2\mathrm{i}  \Omega_s r^2}{n(n+1)} c_n+2\Omega_s r\left[\frac{1}{n}q_n a_{n-1}-\frac{1}{n+1}q_{n+1}a_{n+1}\right] \\-2\Omega_s r^2\left[\frac{n-1}{n}q_n b_{n-1}+\frac{n+2}{n+1}q_{n+1}b_{n+1}\right] \\
  =\nu\left[\frac{1}{r^2}\frac{d}{dr}(r^4\frac{d c_n}{d r})-(n-1)(n+2)c_n \right] \\-2\mathrm{i}\sqrt{\frac{2\pi}{3}}\Omega_s\Omega_p \sin \alpha_s\, r^2\delta_{1n} 
  -2 \mathrm{i} \Omega_s^2 \frac{q_{n+1}}{n+1}\left(r \frac{\mathrm{d} X(r)}{\mathrm{d} r} +3X(r)\right)\delta_{1n} \\+2 \mathrm{i} \Omega_s^2 \frac{q_n}{n}\left(r \frac{\mathrm{d} X(r)}{\mathrm{d} r}-2X(r)\right)\delta_{3n}
\end{multline}
\begin{equation}
 \frac{1}{\rho_0 r^2}\frac{d}{d r}(\rho_0 r^2a_n)-n(n+1)b_n=0.
\end{equation}

% Don't change these lines
\bsp	% typesetting comment
\label{lastpage}
\end{document}